\documentclass[a4paper,12pt]{article}
\usepackage{amssymb,amsthm,amsmath,graphicx,natbib}
\usepackage{algorithm,algorithmicx,algpseudocode,epstopdf,psfrag,epsfig}
\usepackage[textwidth=6.1in,centering]{geometry}
\usepackage[mathlines,running]{lineno}
\newtheorem{proposition}{Proposition}

\def\({\left(}
\def\){\right)}
\def\Var{\text{\rm Var}}

\newcommand{\bq}[1]{\begin{equation}\label{#1}}
\newcommand{\eq}{\end{equation}}
\newcommand{\bqn}{\begin{eqnarray}}
\newcommand{\eqn}{\end{eqnarray}}
\newcommand{\bqns}{\begin{eqnarray*}}
\newcommand{\eqns}{\end{eqnarray*}}

\newcommand{\dd}{\,\textrm{d}}

\newcommand{\Nn}{\textrm{\rm N}}
\newcommand{\wt}{\widetilde}
\newcommand{\diag}{\text{diag}}
\newcommand{\wh}{\widehat}
\newcommand{\mle}{\text{mle}}

\begin{document}

\title{On the existence of moments for high dimensional importance sampling}
\author{Michael K. Pitt$^{(a)}$ \  Minh-Ngoc Tran$^{(b)}$ \\  Marcel Scharth$^{(b)}$ \ Robert Kohn$^{(b)}$ \\
\small{
$^{(a)}$ Economics Department, University of Warwick}\\
\small{
$^{(b)}$  Australian School of Business, University of New South Wales}}

\date{\today}

\maketitle

\vspace{-8mm}
\begin{abstract}
\noindent Theoretical results for importance sampling rely on the existence of certain moments
of the importance weights, which are the ratios between the proposal and target densities.
In particular, a finite variance ensures square root convergence and asymptotic normality of the importance sampling estimate,
and can be important for the reliability of the method in practice.
We derive conditions for the existence of any required moments of the weights for Gaussian proposals
and show that these conditions are almost necessary and sufficient for a wide range of models with latent Gaussian components.
Important examples are time series and panel data models with measurement densities which belong to the exponential family.
We introduce practical and simple methods for checking and imposing the conditions
for the existence of the desired moments.
We develop a two component mixture proposal that allows us to flexibly adapt
a given proposal density into a robust importance density.
These methods are illustrated on a wide range of models
including generalized linear mixed models,
non-Gaussian nonlinear state space models and panel data models with autoregressive random effects.
%\footnotesize{
\bigskip

\noindent {\sc Keywords}:  MCMC, simulated maximum likelihood, state space models, robustness

%\medskip
%\noindent {\sc JEL}: C32, C51, E43.
%}
\end{abstract}

\newpage

\section{Introduction\label{Intro}}
This paper develops robust high-dimensional importance sampling methods for estimating the likelihood of statistical and econometric models with latent Gaussian variables. We propose computationally practical methods for checking and ensuring that the second (and possibly higher) moments of the importance weights are finite. The existence of particular moments of the importance weights is fundamental for establishing the theoretical properties of estimators based on importance sampling. A central limit theorem applies to the importance sampling estimator if the second moment of the weights is finite \citep{Geweke89}. The Berry-Esseen theorem \citep{Berry:1941,Esseen:1942} specifies the rate of convergence to normality when the third moment exists. In a more general case where the weights can have local dependence and may not be identically distributed, if the $n^\text{th}$ moment of the weights exists, \cite{Chen:2004} show that the rate of convergence to normality is of the order of $O(S^{-(n-2)/4})$ with $S$ the number of importance samples. This implies that the higher the moments that exist, the faster the rate of convergence. Even though infinite variance may in some cases be due to a region of the sampling space that has no practical impact for Monte Carlo sampling \citep{oz2001}, this problem can result in highly inaccurate importance sampling estimates in a variety of applications; see \cite{Robert2005} and \cite{ZR2007} for examples.

In many models, e.g. non-Gaussian nonlinear state space models and generalized linear mixed models,
the likelihood involves analytically intractable integrals.
These integrals are often estimated by importance sampling.
A popular approach for estimating the parameters in such models is the simulated maximum likelihood (SML) method \citep[see, e.g.,][]{cGaM1995-2}.
SML first estimates the likelihood by importance sampling
and then maximizes the estimated likelihood.
A second approach for estimating such models is by Bayesian inference using Markov chain Monte Carlo simulation.
If the likelihood is estimated unbiasedly, then the Metropolis-Hastings algorithm
with the likelihood replaced by its unbiased estimator is still able to sample exactly from the posterior.
See, for example, \cite{Andrieu:2009} and \cite{Flury:2011}.
\cite{Pitt:2012b} show that it is crucial for this approach that
the variance of the estimator of the log-likelihood exists,
which is guaranteed by a finite second moment of the weights when the likelihood is estimated by importance sampling.
Both the SML and MCMC methods require an efficient importance sampling estimator of the likelihood.

This paper introduces methods that directly check the existence of any particular moment of the importance weights when using a possibly high-dimensional Gaussian importance density. If the Gaussian importance density does not possess the required moments,
our results make it straightforward to modify the original Gaussian importance density in such a way that the desired moment conditions hold.
This approach is extended by developing a mixture importance density which is a combination of that robust modified Gaussian density with any other importance density. We prove that this mixture importance density satisfies the desired moment conditions, while providing substantial flexibility to design practical and efficient importance densities. The only requirement for the validity of these methods is that the log measurement density is concave or bounded linearly from above as a function of the latent variables. Our method contrasts with previous approaches which rely on statistical testing. See for example  \cite{Monahan(93)} and \cite{Koopman2009}, who develop diagnostic tests for infinite variance based on extreme value theory.

We also develop specific results that allow us to efficiently check and impose the required moment conditions for nonlinear non-Gaussian state space models. Gaussian importance samplers are extensively applied in this setting. Some examples include non-Gaussian unobserved components time series models as in \cite{DK00}, stochastic volatility models in \cite{LR2003}, stochastic conditional intensity models in \cite{BH06}, stochastic conditional duration models in  \cite{BG09}, stochastic copula models in \cite{HafnerManner2008} and dynamic factor models for multivariate counts in \cite{jlr2011}.

We illustrate the new method in a simulation study and in an empirical application to Bayesian inference for a Poisson panel data model with autoregressive random effects.  We consider the \cite{SP97} and \cite{DK97} (SPDK) importance sampling algorithm, a commonly used approach based on local approximation techniques.  We find that the SPDK method leads to infinite variance for this problem. The results show that the mixture approach for imposing the finite variance condition can provide an insurance against poor behavior of the SPDK method for estimating the likelihood at certain parameter values (for example, at the tail of the posterior density), even though the standard method performs nearly as well in most settings. This result suggests that the infinite variance problem typically has low probability of causing instability for this example. We also show that the mixture importance sampler leads to more efficient estimates compared to Student's $t$ importance sampler with the same mean and covariance matrix as the SPDK Gaussian sampler.

Section \ref{sec:IS} provides the background to importance sampling
and establishes the notation. Section \ref{sec:gauss_properties} presents the main theoretical results.
Section \ref{sec:models} discusses the robust importance sampling methods
for three popular classes of models including generalized linear mixed models,
non-Gaussian nonlinear state space models and panel data models.
Section \ref{sec:illustrations} provides several illustrative examples.

%====================================================================%
\section{Importance sampling}\label{sec:IS}
%====================================================================%
In many statistical applications we need to estimate an analytically intractable likelihood of the form
\begin{equation}\label{eq:objective}
L(\psi) = \int p(y|\alpha;\psi)p(\alpha;\psi)d\alpha,
\end{equation}
where $y$ is the observed data, $\alpha$ is a $d\times 1$ vector of latent variables,
and $\psi$ is a fixed parameter vector at which the likelihood is evaluated.
The density $p(y|\alpha;\psi)$ is the conditional density of the data given the latent variable $\alpha$
and $p(\alpha;\psi)$ is the density of $\alpha$.
These densities usually depend on a parameter vector $\psi$.
In some classes of models such as state space models,
\eqref{eq:objective} can be a very high-dimensional integral.
In some other classes such as generalized linear mixed models and panel data models,
the likelihood decomposes into a product of lower dimensional integrals.
Since the parameter vector $\psi$ has no impact on the mathematical arguments that follow, we omit to show this dependence,
except that we still write $L(\psi)$ to explicitly indicate the dependence of the likelihood on $\psi$.

To evaluate the integral \eqref{eq:objective} by importance sampling, we write it as
\begin{equation*} \label{basicid}
    L(\psi)  = \int \frac{p(y|\alpha)p(\alpha)}{g(\alpha)}g(\alpha)\dd\alpha = \int \omega(\alpha) g(\alpha)\dd\alpha,
\end{equation*}
where $g(\alpha)$ is an importance density whose support contains the support of $p(y|\alpha)p(\alpha)$,
and $\omega(\alpha) = {p(y|\alpha)p(\alpha)}/{g(\alpha)}$ is the importance weight.
Let $\alpha_1,\ldots,\alpha_S$ be $S$ draws from the importance density $g(\alpha)$.
Then the likelihood \eqref{eq:objective} is estimated by
$\widehat{L}(\psi)=\frac{1}{S}\sum_{s=1}^{S}\omega(\alpha_s)$.
Let
\begin{equation}\label{eq:n_moment}
E_{g}[\omega(\alpha)^{n}]=\int\omega(\alpha)^{n}g(\alpha)d\alpha.
\end{equation}
If $E_{g}[\omega(\alpha)]<\infty$, a
strong law of large numbers holds, i.e. $\widehat{L}(\psi) \stackrel{a.s.}{\longrightarrow} L(\psi)$
as $S\rightarrow\infty$.
Let $W=\sqrt{S}(\widehat{L}-L)/\sigma_w$ where $\sigma^2_w=\Var_g(w(\alpha))$.
If $E_{g}[\omega(\alpha)^2]<\infty$,
a central limit theorem holds, i.e. $W\stackrel{d}{\to}N(0,1)$ \citep[see, e.g.,][]{Geweke89}.
Furthermore, if $E_{g}[\omega(\alpha)^3]<\infty$, the rate of convergence of $\widehat L$ to normality can be
determined by the Berry-Esseen theorem \citep{Berry:1941,Esseen:1942},
which states that
\[
\sup_z|P(W\leq z)-\Phi(z)|\leq\frac{1}{\sqrt{S}}C\gamma,
\]
where $\gamma=E_{g}[|\omega(\alpha)-L|^3]$, $C$ is a finite constant and $\Phi(z)$ is the standard normal cdf.

Our article first obtains conditions for the existence of the required moments of the weights $\omega(\alpha)$
when using a Gaussian importance density $g(\alpha)=\Nn(\alpha|\mu^{\ast},Q^{\ast-1})$.
The results are then extended to the case of a mixture importance density.

%In practical settings, substantial efficiency gains can be obtained by using
%techniques such as control variables and antithetic variables for variance reduction,
%see for example \cite{Ripley87} and \cite{DK00}. The method consists of generating a set of draws which are perfectly negatively correlated
%with the original draws
%\begin{equation*}
%\alpha_+^{(s)}=2\mu^\ast-\alpha^{(s)},\qquad s=1,\ldots,S,
%\end{equation*}
%and computing the integral estimate as before. This modification has no impact on the theoretical arguments discussed in this paper.

%=============================================================================%
\section{General properties of importance weights\label{sec:gauss_properties}}
%=============================================================================%
We are concerned with the existence of the $n^{th}$ moment $E_g[\omega(\alpha)^n]$ of the importance weights $\omega(\alpha)$
when using a Gaussian importance density $g(\alpha)=\Nn(\alpha|\mu^{\ast},Q^{\ast-1})$.
In this paper, $n$ can be any positive number.
Of special interest is cases where $n$ is a positive integer and we refer to $n=1$ as the fist moment, $n=2$ as the second moment and so forth.
Let $l(\alpha)=\log p(y|\alpha)$. We write the log of the importance weights as
\begin{align*}
\log\omega(\alpha)  &  =l(\alpha)+\log p(\alpha)-\log g(\alpha)\\
&  =c+l(\alpha)+\log p(\alpha)+\frac{1}{2}(\alpha-\mu^{\ast})^{\prime}Q^{\ast
}(\alpha-\mu^{\ast}),
\end{align*}
where the constant $c$ does not depend on $\alpha$.
In this paper, for a square matrix $A$, the notation $A>0$ ($A<0$) means that $A$ is a positive definite (negative definite).
We obtain the following general result on the moments of $\omega(\alpha)$.

\begin{proposition}\label{prop_general}
Suppose that there exists a constant scalar $k$, a vector $\xi$ and a symmetric matrix $Q>0$ such that
\begin{equation}\label{eq:bounding}
l(\alpha)+\log p(\alpha)\leq k-\frac{1}{2}(\alpha-\xi)^{\prime}Q(\alpha
-\xi),\text{ for all }\alpha.
\end{equation}
Then $E_{g}[\omega(\alpha)^{m}]<\infty$ for all $m\leq n$ if $Q^*$ satisfies
\begin{equation}\label{eq:p.d.}
Q^{\ast}-n(Q^{\ast}-Q)>0.
\end{equation}
\end{proposition}

\begin{proof}
See Appendix A.
\end{proof}

When $n = 0$ condition \eqref{eq:p.d.} requires that
the inverse covariance matrix $Q^{\ast}$ of the importance density $g(\alpha)$ is positive definite.
For $n=1$, condition \eqref{eq:p.d.} holds because $Q > 0$ by assumption.
The requirement for a finite variance of the weights,
corresponding to the case $n=2$, requires that $2Q-Q^{\ast}>0$.
In practice the bounding assumption \eqref{eq:bounding} can be difficult to verify for general models.
However, for models in which the density of the latent vector $p(\alpha)$ is Gaussian,
verifying this assumption is straightforward.
Also, when $p(\alpha)$ is Gaussian it is relatively straightforward
to check the existence condition \eqref{eq:p.d.} and to impose conditions on $Q^*$ to ensure the existence of specific moments of the weights.
The rest of this paper will therefore focus on models for which the latent density $p(\alpha)$ is itself multivariate Gaussian $N(\alpha|\mu,Q^{-1})$
with $Q>0$.

\begin{proposition}\label{prop:gauss_trans}
(i) Suppose that there exists a constant scalar $k$, and a vector $\delta$ such that
\begin{equation}\label{eq:linear bound}
l(\alpha)=\log p(y|\alpha)\leq k+\delta'\alpha,\text{ for all }\alpha.
\end{equation}
Then $E_g[\omega(\alpha)^m]<\infty$ for all $m\leq n$ if $Q^{\ast}-n(Q^{\ast}-Q)>0$.
%(ii) Suppose that for at least one $j$ ($j=1,...,d$), there exist constants $\wt k_j$ and $\wt\delta_j$ such that
%\bqn\label{eq:more assumption}
%\text{either}\;\;\lim_{\alpha_j\to-\infty}(l(\alpha)-\wt k_j-\wt\delta_j\alpha_j)=0\;\;\text{or}\;\;\lim_{\alpha_j\to+\infty}(l(\alpha)-\wt k_j-\wt\delta_j\alpha_j)=0.
%\eqn

(ii) Suppose that for at least one $j$ ($j=1,...,d$),
\bqn\label{eq:more assumption}
\text{either}\;\;\lim_{\alpha_j\to-\infty}\frac{l(\alpha)}{\alpha_j^2}=0\;\;\text{or}\;\;\lim_{\alpha_j\to+\infty}\frac{l(\alpha)}{\alpha_j^2}=0.
\eqn
Then $Q^{\ast}-n(Q^{\ast}-Q)<0$ implies $E_g[\omega(\alpha)^n]=\infty$.
\end{proposition}
\begin{proof}
See Appendix A.
\end{proof}
Part (i) of Proposition 2 indicates that
when the log measurement density is concave in the latent variable $\alpha$, a sufficient condition
for the existence of the first $n$ moments is that $Q^{\ast}-n(Q^{\ast}-Q)>0$.
Under the additional assumption of part (ii), the condition is also necessary.
%Proposition \ref{prop:gauss_trans} shows that the positive definiteness of $Q^{\ast}-n(Q^{\ast}-Q)$ is almost necessary and sufficient for the existence of the required moments.
%By ``almost" we mean there is no clear-cut answer to the existence of the $n^\text{th}$ moment if $Q^{\ast}-n(Q^{\ast}-Q)=0$.
%This result follows because $l(\alpha)$ is sub-Gaussian at the tails, so that the existence of moments is entirely governed by the finiteness of the moments of $p(\alpha)/g(\alpha)$.
For many models, it is possible to verify these Assumptions.
%at $l(\alpha)$ is concave or at least bounded linearly from above in $\alpha$.
Assumption \eqref{eq:linear bound} covers
all exponential family models with a canonical link to the latent variable $\alpha$,
i.e. the density $p(y|\alpha)$ is of the form
\begin{equation*}
p(y|\alpha)=\exp\left(\frac{y\eta(\alpha)-b(\eta(\alpha))}{\varpi}\right),
\end{equation*}
where $\eta(\alpha)=z'\alpha+c$, $z$ is a vector of covariates,
$c$ is a constant that does not depend on $\alpha$,
and $\varpi>0$ is a dispersion parameter (see Section \ref{sec:models}).
Because $\ddot{b}(\eta)=\partial^2b(\eta)/\partial\eta^2>0$ by the property of the exponential family,
\begin{equation*}
\frac{\partial^2l(\alpha)}{\partial\alpha\partial\alpha^T}=-\frac{1}{\varpi}\ddot{b}(\eta(\alpha))zz'<0.
\end{equation*}
It follows that $l(\alpha)$ is a concave function in $\alpha$,
because a differentiable function is concave if and only if its Hessian matrix
is negative definite  \cite[see, e.g.,][Chapter 3]{Bazaraa:2006}.
Similarly, we can show that the stochastic volatility model \citep[see, e.g.,][]{Ghysels96}
satisfies \eqref{eq:linear bound} with Gaussian or Student $t$ errors in the observation equation.
The observation equation for the univariate stochastic volatility model with Gaussian errors is
$y_t = \exp(\alpha_t/2)\epsilon_t$, $t=1,...,T$, with $\epsilon_t\sim N(0,1)$.
Hence,
\begin{equation}\label{eq:SV l_alpha}
l(\alpha)=-\frac{T}{2}\log(2\pi)-\frac12\sum_{t=1}^T\alpha_t-\frac12\sum_{t=1}^Ty_t^2\exp(-\alpha_t),\;\;\alpha=(\alpha_1,...,\alpha_T)'.
\end{equation}
It is straightforward to show that the Hessian matrix of $l(\alpha)$ is negative definite, thus $l(\alpha)$ is concave.
The concavity of $l(\alpha)$ in the case with $t$ errors can be shown similarly.

Assumption \eqref{eq:more assumption} is also satisfied by many popular models.
For example, in the Poisson model where $b(\eta(\alpha))=\exp(z'\alpha+c)$
and in the binomial model where $b(\eta(\alpha))=\log(1+e^{z'\alpha+c})$,
we can easily check that \eqref{eq:more assumption} holds.
For the stochastic volatility model with $l(\alpha)$ given in \eqref{eq:SV l_alpha},
\eqref{eq:more assumption} holds because $\lim_{\alpha_j\to+\infty}(l(\alpha)/\alpha_j^2)=0$ for any $j$.

%=============================================================================%
\subsection{A general method for checking and imposing the existence condition}\label{sec:mixture}
%=============================================================================%
This section presents a general method for checking and imposing the existence condition \eqref{eq:p.d.}.
A more efficient method that exploits the structure of non-Gaussian nonlinear state space models is presented in Section \ref{sec:statespace}.

Suppose that $g(\alpha)=\Nn(\alpha|\mu^{\ast},Q^{\ast-1})$ is an importance density available in the literature,
e.g. one obtained by the Laplace method.
It is straightforward to check the positive definiteness of the matrix $Q^{\ast}-n(Q^{\ast}-Q)$
by verifying that all its eigenvalues are positive or by using Sylvester's criterion,
which is discussed in Section \ref{subsecsec:Sylvester}.

If $Q^*$ fails the existence condition,
we modify $Q^*$ to construct a matrix $\wt Q$ such that $\wt Q-n(\wt Q-Q)>0$
by first decomposing $nQ=AA'$ using the Cholesky decomposition.
Let $V$ be the diagonalization matrix of $A^{-1}Q^* {A'}^{-1}$, i.e. $V'A^{-1}Q^* {A'}^{-1}V=\Lambda$,
where $\Lambda=\diag(\lambda_1,...,\lambda_d)$ is a diagonal matrix, $VV'=V'V=I$.
Such an orthonormal matrix $V$ always exists as $A^{-1}Q^* {A'}^{-1}$ is real and symmetric.
We define $\wt\Lambda=\diag(\wt\lambda_1,...,\wt\lambda_d)$ with
\begin{equation}\label{eq:impose1}
\wt\lambda_j=
\begin{cases}
\lambda_j,&\text{if}\;\;\lambda_j<1/(n-1),\\
\frac{1-\epsilon}{n-1},&\text{if}\;\;\lambda_j\geq1/(n-1).
\end{cases}
\end{equation}
for some $\epsilon>0$.
Note that $\wt\lambda_j<1/(n-1)$ for all $j=1,...,d$.
Finally, let $\wt Q=AV\wt\Lambda V'A'$. It follows that $nQ-(n-1)\wt Q=AV(I-(n-1)\wt\Lambda)V'A'>0$.
The new importance density $\wt g(\alpha)=\Nn(\alpha|\mu^{\ast},\wt Q^{-1})$ then satisfies the existence condition \eqref{eq:p.d.}.

For SML, it is essential that the likelihood estimator $\wh L(\psi)$ is continuous,
and desirably, differentiable with respect to the model parameters $\psi$.
The function $\wt\lambda_j=\wt\lambda_j(\lambda_j)$ in \eqref{eq:impose1} is not a continuous function of $\lambda_j$,
so $\wt\lambda_j(\psi)=\wt\lambda_j(\lambda_j(\psi))$ is not continuous in $\psi$.
Appendix C presents a modification of \eqref{eq:impose1} in which $\wt\lambda_j$ is a continuous and differentiable function of $\lambda_j$.

The principle behind the above method is that the difference $\wt Q-Q^*$ is often small,
so that $\wt g(\alpha)$ still provides a good fit to the target density, while the condition for the existence of the $n^{th}$ moment is guaranteed to hold.
Intuitively, only directions of $Q^\ast$ along which the density $N(\alpha|\mu^*,{Q^*}^{-1})$ has light tails
are modified; the other directions are unchanged.
The resulting density $N(\alpha|\mu^*,{\wt Q}^{-1})$ has heavier tails than the original density $N(\alpha|\mu^*,{Q^*}^{-1})$.

However, we have seen that the observation density $p(y|\alpha)$ does not matter for the existence of moments under the assumptions of Proposition \ref{prop:gauss_trans}, so that the importance density $\wt g(\alpha)$
may be a poor approximation to $p(y|\alpha)p(\alpha)$ near its mode. To overcome this problem, we establish the following result.
Let $f(\alpha)=p(y|\alpha)p(\alpha)$ and $\sup(f)=\{\alpha:f(\alpha)\not=0\}$ be the support of $f$.

\begin{proposition}\label{prop_mixture}
Suppose that $g_{1}(\alpha)$ is a density such that $\sup(g_1)\supseteq\sup(f)$,
and for some $n\geq0$,
\bqn\label{eq:moment_g1}
\int_{\sup(g_1)}\left(  \frac{f(\alpha)}{g_{1}(\alpha)}\right)  ^{n}g_{1}(\alpha)d\alpha<\infty.
\eqn
For any density $g_{2}(\alpha)$, consider the mixture importance density
$g(\alpha)=\pi g_{1}(\alpha)+(1-\pi)g_{2}(\alpha)$ with $0<\pi<1$.
Then, for all $m\leq n$,
\[
\int_{\sup(g)}\left(  \frac{f(\alpha)}{g(\alpha)}\right)  ^{m}%
g(\alpha)d\alpha<\infty.
\]
\end{proposition}

\begin{proof}
See Appendix A.
\end{proof}

The proposition suggests an approach to combine the importance density provided by the standard methods with
a second importance density which by itself ensures the existence of the required moments.
This result gives substantial flexibility in that it is useful in practice
to put very little weight on the heavy density component $g_{1}(\alpha)$,
while leaving the rest of the mass for the lighter tailed component.
The existence of the moments will still be entirely governed by the heavier term $g_{1}(\alpha)$,
but the use of the mixture proposal may lead to lower variance since $g_{2}(\alpha)$ is designed to approximate the target density accurately.

In all the models and examples considered in this article, Assumption \eqref{eq:moment_g1} is satisfied with
$g_1(\alpha) = N(\alpha|\mu^*,{\wt Q}^{-1})$
and it is obvious that $\sup(g_1)\supseteq\sup(f)$ as $\sup(g_1)=\mathbb{R}^d$.
The density $g_2(\alpha)$ can be any density that we can sample from.
In practice, we would like to choose $g_2$ such that the weight $\omega(\alpha)$ has a small variance,
while the desired moments are theoretically guaranteed to exist.
In this paper we choose
$g_2(\alpha)=N(\alpha|\mu^*,{Q^*}^{-1})$, and set $\pi=0.1$ based on some experimentation.
We refer to this approach as the $n^\text{th}$-moment constrained mixture importance sampler
or $n^\text{th}$-IS for short.

%=============================================================================%
\section{Models}\label{sec:models}
%=============================================================================%
\subsection{Generalized linear mixed models}\label{Sec:GLMM}
A popular class of models that typically needs importance sampling for likelihood estimation is
generalized linear mixed models (GLMM) \citep[see, e.g.,][]{Jiang:2007}.
Consider a GLMM with data $y_{i}=(y_{i1},...,y_{in_i})'$, $i=1,...,m$.
Conditional on the random effects $\alpha_i$,
the observations $y_{ij}$ are independently and exponentially distributed as
\begin{equation*}
p(y_{ij}|\beta,\alpha_i)=\exp\left(\frac{y_{ij}\eta_{ij}-b(\eta_{ij})}{\varpi}+c(y_{ij},\varpi)\right),
\end{equation*}
with $\eta_{ij}=x_{ij}'\beta+z_{ij}'\alpha_i,\ j=1,...,n_i,\ i=1,...,m,$
where $x_{ij},\ z_{ij}$ are $p$- and $u$-vectors of covariates.
For simplicity, we assume that the dispersion parameter $\varpi$ is known;
the cases with an unknown $\varpi$ require only a small modification to the procedure described below.
The random effects are assumed to have a normal distribution $\alpha_i\stackrel{iid}{\sim}N(0,Q^{-1})$.
The parameters of interest are $\psi=(\beta,Q)$.
The density of $y$ conditional on $\psi$ and $\alpha$ is
\begin{equation*}
p(y|\psi,\alpha)=\prod_{i=1}^mp(y_i|\beta,\alpha_i)=\prod_{i=1}^m\prod_{j=1}^{n_i}p(y_{ij}|\beta,\alpha_i),
\end{equation*}
so that the likelihood is
\begin{equation}\label{e:llh}
L(\psi) = p(y|\psi)=\prod_{i=1}^m L_i(\psi)\;\;\text{with}\;\;L_i(\psi)=\int p(y_i|\beta,\alpha_i)p(\alpha_i|Q)d\alpha_i.
\end{equation}
%which decomposes into a product of lower dimensional integrals.
It is often difficult to estimate the model parameters $\psi$ because the likelihood \eqref{e:llh}
involves analytically intractable integrals $L_i(\psi)$.
A popular approach for estimating $\psi$ is the SML method \citep[see, e.g.,][]{cGaM1995-2}.
SML first estimates the integrals in \eqref{e:llh} by importance sampling
and then maximizes the estimated likelihood over $\psi$.
It is necessary to use common random numbers so that the resulting estimator $\wh L(\psi)$ is smooth in $\psi$.
%\cite{cGaM1995-2} show that the resulting
%estimators are asymptotically (in $S$, the number of importance samples) efficient and asymptotically
%normal provided that $S$ increases at least at rate $m\sqrt{m}$.
Another approach for estimating $\psi$ is by using MCMC
with the likelihood \eqref{e:llh} replaced by its unbiased estimator \citep{Andrieu:2009,Flury:2011,Pitt:2012b}.

Both the SML and MCMC methods require an efficient estimator of the likelihood \eqref{e:llh}.
The robust importance sampling approach for estimating the integrals $L_i(\psi)$
can be carried out as follows.
Write $X_i=[x_{i1},...,x_{in_i}]'$, $Z_i=[z_{i1},...,z_{in_i}]'$, $\eta_i=X_i\beta+Z_i\alpha_i$.
Let $F_i(\alpha_i)=\log p(y_i|\beta,\alpha_i)+\log p(\alpha_i|Q)$, then
\begin{equation*}
F_i(\alpha_i)=\varpi^{-1}\left(y_i'X_i\beta+y_i'Z_i\alpha_i-\mathbf{1}_{n_i}'b(\eta_i)\right)-\frac12\alpha_i'Q\alpha_i+C,
\end{equation*}
where $C$ is the constant term independent of $\alpha_i$. Then
\begin{equation*}
\frac{\partial F_i(\alpha_i)}{\partial \alpha_i}=\varpi^{-1}\left(Z_i'y_i-Z_i'\dot{b}(\eta_i)\right)-Q\alpha_i
\end{equation*}
and
\begin{equation*}
H_i(\alpha_i)=\frac{\partial^2 F_i(\alpha_i)}{\partial \alpha_i\partial \alpha_i'}=-\varpi^{-1}Z_i'\diag(\ddot{b}(\eta_i))Z_i-Q,
\end{equation*}
where $\dot{b}(\eta)$ and $\ddot{b}(\eta)$ are the first and second derivatives of $b(\eta)$
and are understood componentwise, i.e. $\dot{b}(\eta)=(\dot{b}(\eta_1),...,\dot{b}(\eta_k))'$,
$\ddot{b}(\eta)=(\ddot{b}(\eta_1),...,\ddot{b}(\eta_k))'$
when $\eta=(\eta_1,...,\eta_k)'$ is a vector.
It is straightforward to use Newton's method to find a maximizer $\alpha_i^*$ of $F_i(\alpha_i)$,
because both the first and second derivatives of $F_i(\alpha_i)$ are available in closed form.
Let $Q_i^*=-H_i(\alpha_i^*)$.
We can now construct the $n^\text{th}$-moment constrained mixture importance sampler
as described in Section \ref{sec:mixture} to ensure the existence of the first $n$ moments of the weights.
It is therefore straightforward to carry out importance sampling in GLMM with the required moments of the weights
guaranteed to exist.
We note that $l_i(\alpha_i)=\log p(y_i|\beta,\alpha_i)$ is concave in $\alpha_i$ for all GLMMs with the canonical link
(see Section \ref{sec:gauss_properties}),
so that Proposition \ref{prop:gauss_trans}(i) applies.

A common practice is to use a $t$ importance density with mean $\alpha^*_i$,
scale matrix ${Q^*_i}^{-1}$ and small degrees of freedom $\nu$,
although the moments of the weights are not theoretically guaranteed to exist.
Section \ref{Exa:GLMM} compares empirically the performance of such a $t$ importance density
to our constrained mixture importance density.

%-------------------------------------------------%
\subsection{Nonlinear non-Gaussian state space models}\label{sec:statespace}
%-------------------------------------------------%
Consider the following class of nonlinear non-Gaussian state space models
\begin{equation}\label{generaleq}
\begin{split}
y_t|\theta_t\sim p(y_{t}|\theta_t ; \psi),\qquad &\theta_{t}=c_t+Z_t\alpha_{t},\qquad t=1,\ldots,T,\\
\alpha_{t+1}=d+\Phi\alpha_{t}+\eta_{t},\qquad &\alpha_1\sim \Nn(\mu,\Sigma_\alpha), \qquad \eta_{t}\sim \Nn(0,\Sigma),
\end{split}
\end{equation}
where $y_t$ is the $k\times 1$ observation vector, $\theta_t$ is the $p \times 1$ signal vector,
$\alpha_t$ is the $m\times1$ state vector, and $Z_t$ is the $p\times m$ selection matrix;
the $m\times 1$ constant vector $d$, the $m\times m$ transition matrix $\Phi$ and the $m\times m$ covariance variance matrix $\Sigma$
jointly determine  the dynamic properties of the model.
%The system matrices $Z_t$, $G_t$ and $H_t$ are time-varying and non-stochastic.
The parameter vector $\psi$ contains the unknown coefficients in the
observation density and in the system matrices. A thorough overview of the models
and methodology is provided in \cite{DK1}.

Define $\alpha=(\alpha_1' \, , \, \ldots \, , \, \alpha_T')'$ and $y=(y_1' \, , \, \ldots \, , \, y_T')'$. The likelihood for the model is
\begin{equation}\label{eq:lik}
L(\psi)=\int p(y|\alpha)p(\alpha)\dd\alpha=
\int \prod_{t=1}^{T} p(y_t|\alpha_t)p(\alpha_1)\prod_{t=2}^{T}p(\alpha _t|\alpha_{t-1})\dd\alpha_1\ldots \dd\alpha_T,
\end{equation}
with $p(\alpha)=p(\alpha_1)\prod_{t=2}^{T}p(\alpha _t|\alpha_{t-1})$ and $p(y|\alpha)=\prod_{t=1}^{T}p(y_t|\alpha_{t})$. The latent states are Gaussian, with $p(\alpha)=N(\alpha|\mu;Q^{-1})$,
where $Q$ is a block tridiagonal matrix because the states $\alpha_t$
follow a first order autoregressive process.
A closed form representation of $Q$ is derived in Appendix D.

A suitable Gaussian importance density for evaluating the likelihood \eqref{eq:lik} is
$g(\alpha) = g(y|\alpha)p(\alpha)/g(y)$,
where $g(y)=\int g(y|\alpha)p(\alpha)d\alpha$ and
\begin{equation}\label{idensity}
\begin{split}
g(y|\alpha) = \prod _{t=1}^T g(y_t|\alpha _t),\;\;\;\;g(y_t |\alpha_t) = \exp \left\{a_t + b_t^{\prime} \, \alpha_t-\frac 12 \alpha_t^{\prime}\, C_t \, \alpha_t \right\}.
\end{split}
\end{equation}
Let $B=(b_1',\ldots \,b_T')'$, $C=\textrm{diag}(C_1,\ldots,C_T)$.
\cite{SP97} and \cite{DK97} propose an efficient method for selecting the importance parameters $B$ and $C$,
which is referred to as the SPDK method. Appendix B provides the details. 
We apply the SPDK method in Section \ref{sec:illustrations}. 
It is simple to verify that $g(\alpha)$ is a multivariate normal density with inverse covariance
$Q^*=C+Q$ and mean $\mu^*=Q^{\ast-1}(B+Q\mu)$.

\cite{SP97} and \cite{DK97} show that we can efficiently sample from $g(\alpha)$ and calculate its integration constant by interpreting $g(\alpha)$ as an approximating linear Gaussian state space model with observations $\widehat{y}_t= C_t^{-1}b_t$ and the linear Gaussian measurement equation
\begin{equation}\label{eq:apss}
\begin{split}
    &\widehat{y}_t =\alpha_{t}+\varepsilon_{t},\qquad \varepsilon_{t}\sim \Nn(0,C_t^{-1}), \qquad t=1,\ldots,T.\\
\end{split}
\end{equation}
The simulation smoothing methods of \cite{DJS1995} and \cite{DK02} sample $\alpha$ from $g(\alpha|\widehat{y})$ in $O(T)$ operations. The Kalman filter calculates the integration constant $g(\widehat{y})$ by evaluating the likelihood function for the linear state space model \eqref{eq:apss}. \cite{JK2007} show that it is only necessary that the individual matrices $C_t$ are non-singular for the Kalman filter based procedure to be valid.

%------------------------------------------------------------%
\subsubsection{Checking the existence of moments}\label{subsecsec:Sylvester}
%------------------------------------------------------------%

If the first $n$ moments are required, then by Proposition \ref{prop:gauss_trans} it is only necessary to check that $Q^*-n(Q^*-Q)=Q-(n-1)C>0$, with $Q^*=C+Q$. It is possible to exploit the special structure of the non-Gaussian state space model
to obtain a fast $O(T)$ method for checking and imposing this condition. This method is more computationally efficient than the eigenvalue method described in Section \ref{sec:mixture} when $T$ is large.

Suppose now that $\alpha_t$ is scalar.
The main principles and procedures in this section
are best illustrated and analytically explored for the case with scalar $\alpha_t$.
We discuss the case with multivariate $\alpha_t$ below.
The stationary univariate AR(1) model for $\alpha_t$ is
\bqn\label{eq:ar1_state}
\alpha_{t+1}=\mu(1-\phi)+\phi\alpha_{t}+\eta_{t},\;\;\;\;\eta_t\sim N(0,\sigma^2),
\eqn
for $t=1,...,T-1$ with $\left\vert \phi\right\vert <1$ and initial condition
$\alpha_{1}\sim N(\mu,\sigma_{\alpha}^{2})$, where the unconditional variance
of the states is $\sigma_{\alpha}^{2}=\sigma^{2}/(1-\phi^{2})$.
The observation equation for the approximating linear state space model \eqref{eq:apss} is
\begin{equation}\label{eq:pseudo_meas}
\begin{split}
    &\widehat{y}_t =\alpha_{t}+\varepsilon_{t},\qquad \varepsilon_{t}\sim \Nn(0,v_t), \qquad t=1,\ldots,T,\\
\end{split}
\end{equation}
with $v_t=C_t^{-1}$.
The matrix $Q-(n-1)C$ has the tridiagonal form (see Appendix D)
\begin{equation}\label{eq:matrix_form}
\sigma^{-2}
\begin{pmatrix}
1-\frac{\sigma^{2}(n-1)}{v_{1}}& -\phi & \ldots & 0\\
-\phi & 1+\phi^{2}-\frac{\sigma^{2}(n-1)}{v_{2}}& -\phi & \vdots\\
\vdots & -\phi & 1+\phi^{2}-\frac{\sigma^{2}(n-1)}{v_{T-1}} & -\phi\\
0 & \cdots & -\phi & 1-\frac{\sigma^{2}(n-1)}{v_{T}}
\end{pmatrix}.
\end{equation}
To check that this matrix is positive definite,
we use Sylvester's criterion \citep[see, e.g.,][Chapter 7]{Horn:1990} which states that a symmetric real matrix is positive definite
if and only if all of the leading principal minors are positive. This means
that all the determinants of the square upper left sub-matrices, in our case,
of the $T\times T$ matrix $Q-(n-1)C$ must be positive. Denote the
determinant of the top left $1\times1$ sub-matrix as $\Lambda
_{1}=1-\sigma^{2}(n-1)v^{-1}_{1}$. Setting $\Lambda_{0}=1$, it can be checked that the top
left $t\times t$ sub-matrix has determinant $\Lambda_{t}$ which satisfies the recursion
\begin{equation}\label{eq:diff eq1}
\Lambda_{t}=\{1+\phi^{2}-\sigma^{2}(n-1)v^{-1}_{t}\}\Lambda_{t-1}-\phi^{2}%
\Lambda_{t-2},
\end{equation}
for $t=2,..,T-1$ and%
\begin{equation}\label{eq:diff eq2}
\wt\Lambda_{T}=\{1-\sigma^{2}(n-1)v^{-1}_{T}\}\Lambda_{T-1}-\phi^{2}\Lambda_{T-2}.
\end{equation}
It is therefore necessary that all the determinants $\Lambda_{t},\ t=1,...,T-1$ and $\wt\Lambda_{T}$
are positive for the matrix $Q-(n-1)C$ to be positive definite,
which gives a straightforward and fast approach for checking whether the required
moments exist.

%--------------------------------------------%
\subsubsection{Imposing the existence of moments\label{sec:impose}}
%--------------------------------------------%
We now examine a simple condition on $v_{t}$ which ensures that the first $n$ moments exist.
%For nonlinear non-Gaussian state space models,
%this method is more computationally efficient than the general method introduced in Section \ref{sec:mixture}.
%The reason is that general method leads to a new matrix of importance parameters $\wt C$ which is not diagonal,
%which breaks down the approximating linear state space model structure of the sampler
%and precludes us from using the SPDK method for efficiently sampling from $g(\alpha)$.
%With a general covariance matrix for the proposal, the computational cost of generating importance samples is $O(T^2)$.
%can use the method described in Section \ref{sec:mixture}
%to check and impose the existence condition that $Q-(n-1)C>0$.
%However, the resulting modified matrix $\wt Q$ may not have the block diagonal structure,
%which precludes us from using the SPDK method for efficiently sampling from $g(\alpha)$.
%However, this general method is expensive when $T$ is large.
We want to find a constant value for $v=v_{t}$
in \eqref{eq:pseudo_meas} for which the resulting matrix $Q-(n-1)C_v$ is positive definite,
with $C_v=\diag(1/v,...,1/v)$ .
This corresponds to a constant variance in
the measurement density of the approximating state space form,
\begin{equation}\label{eq:pseudo_meas1}
\widehat{y}_{t}=\alpha_{t}+\varepsilon_{t},\text{ }\varepsilon_{t}\sim
N(0,v).
\end{equation}
Proposition \ref{prop:pseudo_var} shows that $Q-(n-1)C_v>0$ if $v=(n-1)\sigma_{\alpha}^{2}\frac{1+|\phi|}{1-|\phi|}$ when
$\phi\not=0$ and $v=(n-1)\sigma_{\alpha}^{2}+\epsilon$ when $\phi=0$, for some small $\varepsilon>0$.
We set $\varepsilon=10^{-5}$ in this paper.

\begin{proposition}\label{prop:pseudo_var}
Consider the non-Gaussian nonlinear state space model given in \eqref{generaleq} with scalar $\alpha_t$.
Suppose there exists a constant scalar $k$ and a vector $\delta$ such that
\[
l(\alpha_{t})=\log p(y_t|\alpha_{t})\leq k+\delta'\alpha_{t},\text{ for all
}\alpha_{t}.
\]
Suppose that the proposal density is $g(\alpha)=N(\alpha|\mu^{\ast},Q^{\ast-1})$ with $Q^*=Q+C_v$.
Then $E_g[\omega(\alpha)^n]<\infty$ provided that
\begin{equation}\label{eq:cond_ss}
v\geq(n-1)\sigma_{\alpha}^{2}\frac{1+|\phi|}{1-|\phi|}\;\;\text{if}\;\;\phi\not=0,\;\;\text{and}\;\; v>(n-1)\sigma_{\alpha}^{2}\;\;\text{if}\;\;\phi=0.
\end{equation}
\end{proposition}
\begin{proof}
See Appendix A.
\end{proof}

\begin{figure}[h]
\centering
\includegraphics[width=.8\textwidth,height =.4\textheight]{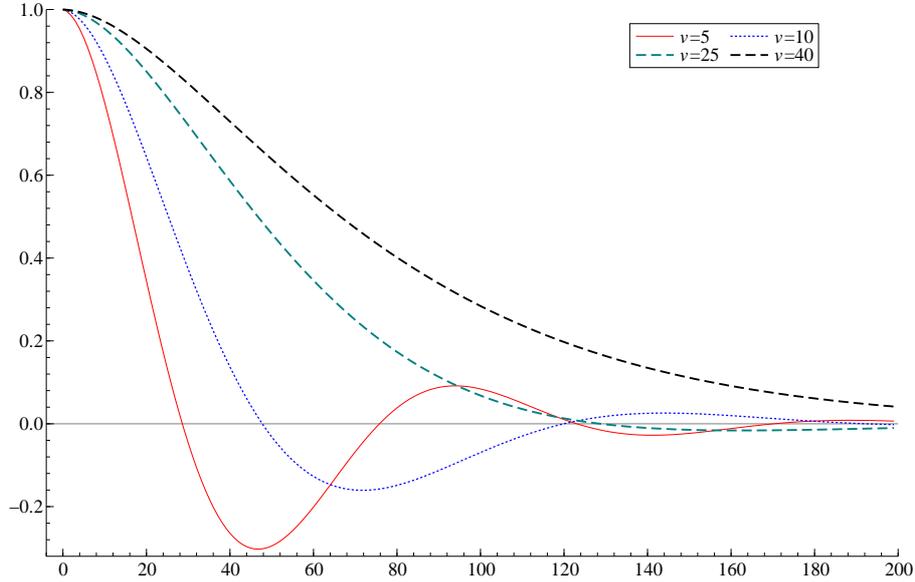}
\caption{
Plots of the determinants $\Lambda_{t}$ against $t$ for $\phi=0.975$, $\sigma_{\alpha}^{2}=0.5$
and four values of the approximating measurement variance $v$.
Under Assumption \eqref{eq:more assumption},
any negative values in the plot show that
the second moment of the weights does not exist.
}
\label{fig:deter}
\end{figure}

Figure \ref{fig:deter} plots the determinants $\Lambda_t$ against $t$
when $n=2$, $\phi=0.975$, and $\sigma_{\alpha}^{2}=0.5$ for $v=5$, 10, 25 and 40.
Since $v\geq\sigma_{\alpha}^{2}(1+|\phi|)/(1-|\phi|)=39.5$, $\Lambda_t$ decays exponentially only for $v=40$, with the three smaller values of $v$ resulting in complex roots
in the difference equation \eqref{eq:diff eq1}, 
and therefore sinusoidal paths that cross the horizontal axis.
If the model is less persistent, i.e. $\phi$ is smaller, then
it is unnecessary for $v$ to be as large. For example, suppose that $n=2$, with $\sigma_{\alpha}^{2}=0.5$ as before and $\phi=0.6$. Then it is only necessary that $v\geq2$ for the second moment to be finite.

Using the approximating state space model \eqref{eq:pseudo_meas1} with constant variance $v$
as the importance density may be inefficient because it does not depend on the $v_t$
which we estimate from the data.
To take greater account of the computed $v_t$, let $v_t^{(0)}=v_t$,
$C^{(0)}=\diag(1/v_1^{(0)},...,1/v_T^{(0)})$, and $\epsilon>0$ be some small value.
For $k=1,2,...$, define $v_t^{(k)}=v_t^{(k-1)}$ if $v_t^{(k-1)}\geq v$ and $v_t^{(k)}=v_t^{(k-1)}(1+\epsilon)$ if $v_t^{(k-1)}<v$.
Let $C^{(k)}=\diag(1/v_1^{(k)},...,1/v_T^{(k)})$.
We suggest the following algorithm for finding the smallest $k\geq 0$ such that $Q-(n-1)C^{(k)}>0$.

\noindent{\bf Algorithm 1.} Set $k=0$.
\begin{enumerate}
\item Check the positive definiteness of $Q-(n-1)C^{(k)}$ as in Section \ref{subsecsec:Sylvester}.
\item If it is positive definite then stop. Otherwise, set $k=k+1$ and go back to Step 1.
\end{enumerate}
It is straightforward to see that the algorithm always converges,
as with a large enough $k$, all the $v_t^{(k)}\geq v$ and therefore
$Q-(n-1)C^{(k)} = Q-(n-1)C_v+(n-1)(C_v-C^{(k)})>0$.
%Indeed, consider those $\gamma_k$ for which $\gamma_k\leq\min_t(v_t)/v$ and note that $v_t^{(k)}\geq v_t^{(k-1)}\geq...\geq v_t$.
%If $v_t^{(k-1)}\geq v$ then $v_t^{(k)}=v_t^{(k-1)}\geq v$, if $v_t^{(k-1)}< v$ then $v_t^{(k)}=v_t^{(k-1)}/\gamma_k\geq v_t/\gamma_k\geq v$.
%This implies that $C_v-C^{(k)}\geq 0$,
%and therefore $Q-(n-1)C^{(k)} = Q-(n-1)C_v+(n-1)(C_v-C^{(k)})>0$.
This justification does not mean that all $v_t^{(k)}\geq v$ after convergence.
The motivation for Algorithm 1 is to adjust the data-based variances $v_t$ as little as possible
while ensuring the existence of the required moments.

After the algorithm converges, let $C^*=C^{(k)}$ to obtain $Q-(n-1)C^*>0$.
In practice, it is more efficient to use a mixture of two
approximating Gaussian models in the form of \eqref{eq:pseudo_meas},
one with $\varepsilon_{t}\sim \Nn(0,v_t^*)$ and the other with $\varepsilon_{t}\sim \Nn(0,v_t)$,
as justified by Proposition \ref{prop_mixture}.
We put a small weight $\pi=0.1$ on the former in all the examples below.

%------------------------------------------------------------%
\subsubsection{Checking and imposing the existence of moments for multivariate $\alpha_t$}\label{subsecsec:multi alpha}
%------------------------------------------------------------%
Let $C^{(k)}=C^{(k-1)}/(1+\epsilon)$ for $k=1,2,...$ with $C^{(0)}=C$ and $\epsilon>0$ is a small number.
The following algorithm provides a way to find the smallest $k$ such that $Q-(n-1)C^{(k)}>0$.

\noindent{\bf Algorithm 2.} Set $k=0$.
\begin{enumerate}
\item Compute the smallest eigenvalue $\lambda_{\min}$ of $Q-(n-1)C^{(k)}$.
\item If $\lambda_{\min}>0$ then stop. Otherwise, set $k=k+1$ and go back to Step 1.
\end{enumerate}
After convergence, as a by-product for checking the existence condition,
a resulting $k>0$ means that the original $C$ fails the condition.
Computing $\lambda_{\min}$ will be fast because $Q-(n-1)C^{(k)}$ is symmetric and block tridiagonal \citep{Saad:2011}.
We now construct a mixture of two approximating Gaussian models \eqref{eq:apss}
using the matrices $C$ and $C^*=C^{(k)}$.

%When $y_t$ is scalar, the signal $\theta_t$ is also scalar for most of the popular models,
%so that working with $\theta=(\theta_1,...,\theta_T)'$ as the latent variable
%will be sometimes more computationally efficient than working with original $\alpha$ \citep{DK97}.
%The density of $\theta$ then is a multivariate normal density with covariance $Q_\theta^{-1}=ZQ^{-1}Z'$,
%where $Z$ is a $T\times mT$ block diagonal matrix $\diag(Z_t,t=1,...,T)$.
%Note that $Q_\theta$ is a block tridiagonal matrix.

%The model is
%\bqn\label{eq:ar1_state}
%y_{t} \sim p(y_{t}|\alpha_{t}),\;\;\;\alpha_{t+1}=\mu(1-\phi)+\phi\alpha_{t}+\sigma u_{t},
%\eqn
%for $t=1,...,T$ with $\left\vert \phi\right\vert <1$ and initial condition
%$\alpha_{1}\sim N(\mu;\sigma_{\alpha}^{2})$, where the unconditional variance
%of the states is $\sigma_{\alpha}^{2}=\sigma^{2}/(1-\phi^{2})$. In this case, the observation density for the approximating linear state space model \eqref{eq:apss} becomes simply
%\begin{equation}\label{eq:pseudo_meas}
%\begin{split}
%    &\widehat{y}_t =\alpha_{t}+\varepsilon_{t},\qquad \varepsilon_{t}\sim \Nn(0,v_t), \qquad t=1,\ldots,T\\
%\end{split}
%\end{equation}
%with $v_t=C_t^{-1}$. See Appendix B for the estimation of the importance parameters $B$ and $C$.

%-------------------------------------------------%
\subsection{Panel data models with an AR(1) latent process}\label{sec:panel}
%-------------------------------------------------%
The random effects models considered in Section \ref{Sec:GLMM} do not take into account time-varying
individual effects. A possible way to overcome this is to include in the model a time-varying individual-specific effect $\alpha_{it}$.
Let $y_i=(y_{i1},...,y_{iT_i})'$, $i=1,...,m$, be panel data,
which are modeled as
\begin{equation}\label{e:panel_AR1}
p(y_{it}|\beta,\alpha_i)=\exp\left(\frac{y_{it}\eta_{it}-b(\eta_{it})}{\varpi}+c(y_{it},\varpi)\right),
\end{equation}
with $\eta_{it}=x_{it}'\beta+\alpha_{it},\ t=1,...,T_i,\ i=1,...,m$,
where the $x_{it}$ are $p$-vectors of covariates.
The random effects $\alpha_i=(\alpha_{i1},...,\alpha_{iT_i})'$ are assumed to follow the AR(1) process
\begin{equation*}
\alpha_{i,t+1}=\mu(1-\phi)+\phi\alpha_{it}+\sigma u_{it},\;\;t=1,...,T_i,
\end{equation*}
with $|\phi|<1$ and initial value $\alpha_{i1}\sim N(\mu,\sigma^2/(1-\phi^2))$, $u_{it}\stackrel{iid}{\sim}N(0,1)$.
The parameter vector $\psi$ consists of $\beta$, $\phi$ and $\sigma^2$.
See, e.g., \cite{Bartolucci:2012}.

The likelihood is
\begin{equation}\label{eq:likelihood_panel}
L(\psi) = p(y|\psi)=\prod_{i=1}^mp(y_i|\psi)=\prod_{i=1}^m\int p(y_i|\alpha_i,\psi)p(\alpha_i|\psi)d\alpha_i,
\end{equation}
which decomposes into a product of lower dimensional integrals, where
\begin{equation*}
p(y_i|\alpha_i,\psi) = \prod_{t=1}^{T_i}p(y_{it}|\beta,\alpha_{it})\,\;\;\text{and}\;\;p(\alpha_i|\psi)=p(\alpha_{i1})\prod_{t=1}^{T_i-1}p(\alpha_{i,t+1}|\alpha_{it}).
\end{equation*}
Each panel $y_i$ has the form of a non-Gaussian non-linear state space model \eqref{eq:poisson}
except that $T_i$ is often small,
therefore the SPDK method can be used to obtain the importance parameters $\mu^*_i$ and ${Q^*_i}^{-1}$.
Because the lengths $T_i$ of the panels are often small,
we can use the general method in Section \ref{sec:mixture} to impose the condition for the existence of moments.

%==========================================================%
\section{Illustrations}\label{sec:illustrations}
%==========================================================%

\subsection{Illustrative example}\label{subsec:toy example}
Let $y = (y_1,...,y_N)$ be observations from a Bernoulli distribution with
success probability $\alpha\in(0,1)$.
The likelihood is
\begin{equation*}
p(y|\alpha)=\prod_{i=1}^N\alpha^{y_i}(1-\alpha)^{1-y_i}=\alpha^k(1-\alpha)^{N-k},
\end{equation*}
with $k$ the total number of successes.
We use a normal distribution $N(0.5,Q^{-1})$, with $Q=0.1$, truncated to the interval $(0,1)$
as a relatively uninformative prior on $\alpha$.
That is, $p(\alpha)=C\kappa(\alpha)1_{(0,1)}(\alpha)$, where $\kappa(\alpha) =  \exp\left(-\frac12Q(\alpha-0.5)^2\right)$,
with $C$ the normalizing constant such that $\int_0^1p(\alpha)d\alpha=1$.
Suppose that we are interested in estimating the posterior mean of $\alpha$,
i.e. we wish to estimate the integral $I = \int_0^1\alpha p(\alpha|y)d\alpha$
using importance sampling.
Let $g(\alpha)$ be the importance density. Then the unnormalized weight is given by
$w(\alpha) = {1_{(0,1)}(\alpha)\kappa(\alpha)p(y|\alpha)}/{g(\alpha)}$.
It is possible to use the Laplace method to construct a Gaussian importance density,
but the constraint on the interval $(0,1)$ makes it difficult to find the mode of $p(\alpha|y)\propto p(\alpha)p(y|\alpha)$
when the mode is near either boundary.
We instead use a second order Taylor expansion of $l(\alpha)=\log p(y|\alpha)$ at the MLE estimate $\wh\alpha_\mle=k/N$ of $\alpha$.
Then, for all $\alpha\in(0,1)$,
\begin{eqnarray*}
\log p(y|\alpha)+\log p(\alpha)&=&\text{constant}+l(\alpha)-\frac12Q(\alpha-0.5)^2\\
&\simeq&\text{constant}+l(\wh\alpha_\mle)+\frac12D(\alpha-\wh\alpha_\mle)^2-\frac12Q(\alpha-0.5)^2\\
&=&\text{constant}-\frac12Q^*(\alpha-\alpha^*)^2,
\end{eqnarray*}
where
\begin{equation*}
D=\frac{\partial^2 l(\alpha)}{\partial\alpha^2}\big|_{\alpha=\wh\alpha_\mle}=-\frac{k}{\wh\alpha_\mle^2}-\frac{N-k}{(1-\wh\alpha_\mle)^2},\;\;
Q^*=Q-D,\;\;\alpha^*=\frac{0.5Q-D\wh\alpha_\mle}{Q^*}.
\end{equation*}
Suppose that $g(\alpha) = N(\alpha|\alpha^*,{Q^*}^{-1})$ is the importance density. Then the second moment of the unnormalized weight $E_g[w(\alpha)^2]$
is approximated as
\begin{equation*}
E_g[w(\alpha)^2] \simeq K\int_0^1\exp\left(-\frac12(2Q-Q^*)(\alpha-\wt\alpha)^2\right)d\alpha,
\end{equation*}
with $\wt\alpha=({Q-Q^*\alpha^*})/({2Q-Q^*})$ and a finite constant $K$.
In theory, unlike the result in Proposition \ref{prop_general},
this second moment exists even if $2Q-Q^*=Q+D<0$,
because the integral is taken over the interval $(0,1)$ rather than the real line as in \eqref{eq:n_moment}.
However, in practice, this integral can be very large if $2Q-Q^*=Q+D\ll0$.
This is likely to happen as $D$ is often very small, especially in extreme cases where the underlying $\alpha$ is close to 0 or 1,
while we need to keep $Q$ small enough in order to have a flat prior.
We refer to this as the normal-IS case.

To demonstrate the theory presented in Section \ref{sec:gauss_properties},
we use a $2^\text{nd}$-moment constrained mixture importance sampler $g(\alpha)=\pi g_1(\alpha)+(1-\pi)g_2(\alpha)$ constructed as in Section \ref{sec:mixture}.
The heavy-tailed component $g_1(\alpha)$ with the weight $\pi=0.1$
is a normal density $N(\alpha^*,{\wt Q}^{-1})$ with $\wt Q$ modified from $Q^*$ such that $2Q-\wt Q>0$,
and $g_2(\alpha)$ is the original normal density $N(\alpha^*,{Q^*}^{-1})$.
We refer to this case as the $2^\text{nd}$-IS. Using a $t$ importance density rather than a normal importance density
is sometimes recommended in the literature \citep{Geweke89}.
As the third importance density, we use a $t$ density with location $\alpha^*$, scale ${Q^*}^{-1}$
and small degrees of freedom $\nu$.
We set $\nu=5$ in this example after some experimentation and refer to this as the $t$-IS case.

\begin{figure}[h]
\centering
\includegraphics[width=1\textwidth,height =.5\textheight]{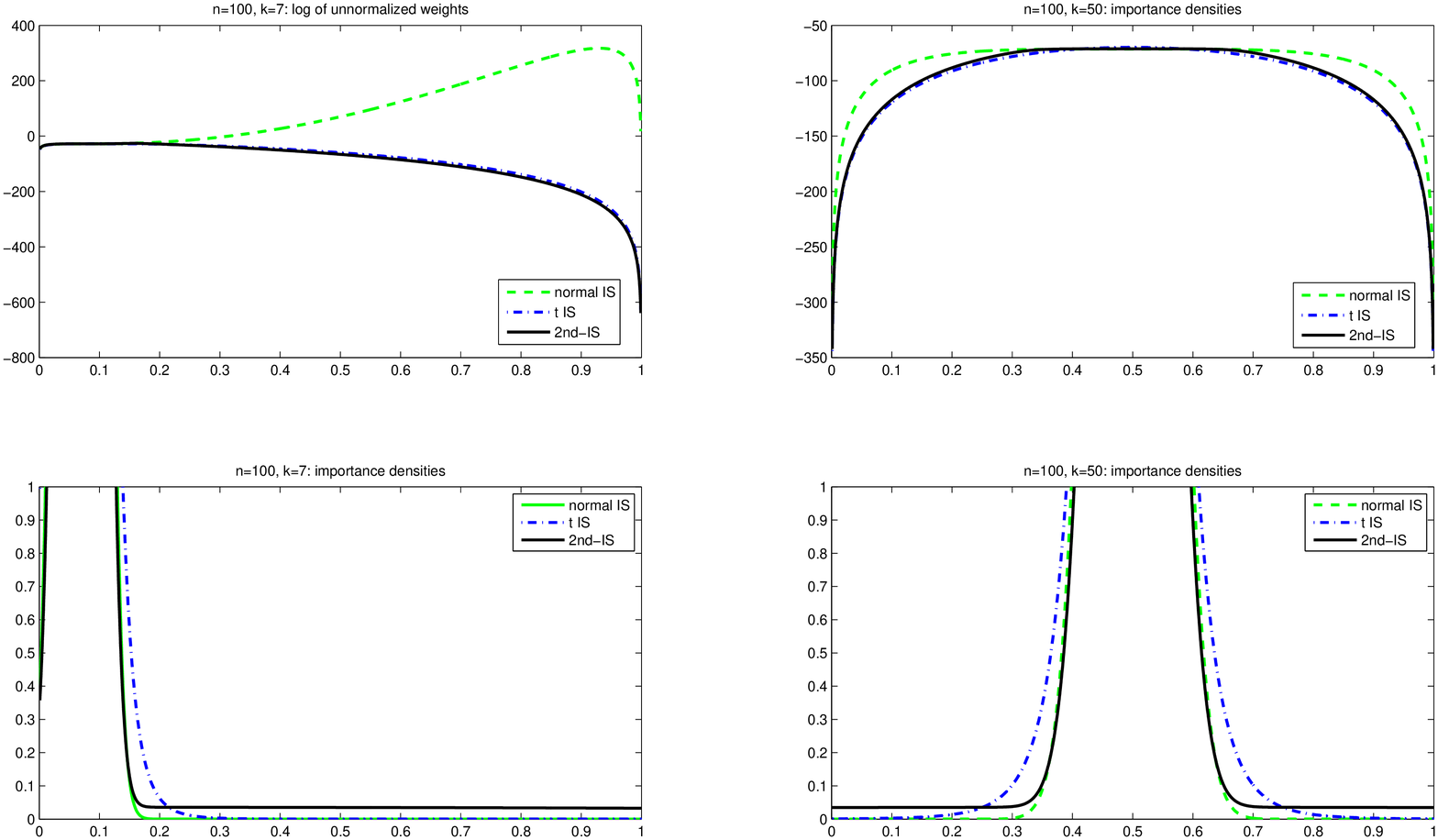}
\caption{Bernoulli example: plots of the log weight functions $\log(w(\alpha))$ against $\alpha$
for three importance sampling approaches.
}
\label{f:toy_example}
\end{figure}

The first row of Figure \ref{f:toy_example} plots $\log(w(\alpha))$ against $\alpha$ for the three importance densities described above
for two cases: a `hard' case  where $N=100$, $k=7$ and an `easy' case where $N=100$, $k=50$.
In the hard case the weight function $w(\alpha)$ of the normal-IS behaves badly for large $\alpha$,
because the posterior is highly skewed to the right and the tail of the normal importance density declines much faster than the posterior.
The figure shows that $t$-IS and $2^\text{nd}$-IS both avoid this problem.
In the easy case, where the posterior is symmetric and is well approximated by a normal distribution,
all the three weight functions appear well behaved.
The panels in the second row plot the three importance densities,
which show that imposing the condition leads to an importance density with the necessary heavy tails.

Let $\alpha_s\stackrel{iid}{\sim}g(\alpha)$, $s=1,...,S=10^6$ be $S$ samples generated from the importance density $g(\alpha)$.
The IS estimator of the integral $I$ and its asymptotic variance are \citep{Geweke89}
\begin{equation*}
\wh I = \frac{\sum_{s=1}^S\alpha_sw(\alpha_s)}{\sum_{s=1}^Sw(\alpha_s)},\;\;\;\;\wh\Var(\wh I) = \frac{S\sum_{s=1}^S(\alpha_s-\wh I)^2w(\alpha_s)^2}{\big(\sum_{s=1}^Sw(\alpha_s)\big)^2}.
\end{equation*}

We implemented the statistical test for finite variance by \cite{Koopman2009} (KSC).
Define a sequence of $r$ random variables $Z_1,\ldots,Z_r$  created from the importance weights
that exceed a given threshold $u$ as $Z_i=\omega(\alpha_i)-u$.
The threshold $u$ is set to the $90^\text{th}$ percentile.
Their test is based on the fact as $S$ and $u$ increase,
the distribution of the excesses $Z_1,\ldots,Z_r$ converges to the generalized Pareto distribution
\[f(z;\xi,\beta)=\frac1\beta\left(1+\xi\frac{z}{\beta}\right)^{-\frac1\xi-1}.\]
The shape parameter $\xi$ characterizes the thickness of the tails of the distribution.
It can be shown that when $\xi>0$, $E_g[\omega(\alpha)^n]=\infty$ for $n\geq1/\xi$.
Hence, the variance of the importance weights is infinite when $\xi>1/2$.
The \cite{Koopman2009} tests consists of obtaining a maximum likelihood estimate of $\xi$ and testing the null hypothesis that $\xi\leq1/2$.
We implement the Wald test version of their procedure
and reject the hypothesis that the variance is finite if the returning $p$-value is smaller than 0.01.

Table \ref{tab:toy_example} reports the following items averaged over 100 replications:
(i) the estimate of the posterior mean $\wh I$,
(ii) the ratio of the asymptotic variance of $\wh I$ for each IS approach
divided by the asymptotic variance for the normal-IS,
(iii) the proportion of the 100 replications in which the KSC test rejects the hypothesis of a finite variance,
(iv) and the CPU time in seconds.
In the ``hard" case $N=100$ and $k=7$, the asymptotic variances of the $t$-IS and $2^\text{nd}$-IS
are both small compared to that of the normal-IS,
and the KSC test almost always rejects the hypothesis of a finite variance of the normal-IS.
The improvement of the $2^\text{nd}$-IS over the normal-IS and $t$-IS
is bigger when the ratio $k/N$ is closer to either 0 or 1.
In the ``easy" case $N=100$ and $k=50$,
the normal IS has the smallest asymptotic variance,
the KSC test indicates that all the variances are finite.
However, we note that the loss in efficiency when using the $2^\text{nd}$-IS in the ``easy" case
is not as severe as the loss for the normal-IS in the ``hard" case.
That is, in general the $2^\text{nd}$-IS should be used
if we do not have much information about the target density.

We found that the KSC test is very sensitive to the selection of the threshold $u$.
In the examples follow we therefore do not compare our results to the test.

\begin{table}[htbp]
\begin{center}
\begin{tabular}{l|ccc|ccc}
\hline
& \multicolumn{3}{c|}{$N=100,k=7$} & \multicolumn{3}{c}{$N=100,k=50$}\\
\hline
Importance density	&normal&$t$&$2^\text{nd}$-IS&normal	&$t$	&$2^\text{nd}$-IS\\
$\wh I$			&0.078	&0.078	&0.078		&0.5	&0.5	&0.5	\\
Variance ratio		&1	&0.38	&0.34		&1	&1.83	&1.20	\\
KSC rejections		&99	&0	&0		&0     	&0	&0	\\
CPU time		&0.24	&0.39	&0.35		&0.26	&0.39	&0.37\\
\hline
\end{tabular}
\end{center}
\caption{Bernoulli example: The table reports the estimate of the posterior mean,
the ratio of the asymptotic variances, the KSC test rejections and the CPU times for the normal-IS, the $t$-IS and the $2^\text{nd}$-IS.
All the values are averaged over 100 replications.}
\label{tab:toy_example}
\end{table}

%==========================================================%
\subsection{Likelihood evaluation for a non-Gaussian non-linear state space model}\label{subsec:non-Gau_ss}
%==========================================================%
We consider a time series $\{y_t, t=1,...,T\}$ generated from the dynamic Poisson model
\begin{eqnarray}\label{eq:poisson}
p(y_{t}|\lambda_{t})&=&\frac{\exp(-\lambda_{t})\lambda_{t}^{y_{t}}}{y_{t}!},\;\;\log(\lambda_{t})=\beta+\alpha_{t},\\
\alpha_{t+1}&=&\phi \alpha_{t}+\eta_{t},\;\; \eta_{t}\sim N(0,\sigma^2),\;\; \alpha_{1}\sim N(0,\sigma^2/(1-\phi^2))\notag.
\end{eqnarray}
We generate the data using the parameter values $\beta=-1.4$, $\phi=0.8$ and $\sigma^2=0.5(1-\phi^2)$.
The objective is to determine whether the importance weights from the SPDK importance sampling method
for estimating the likelihood in model \eqref{eq:poisson}
have a finite variance.
We also investigate the potential benefits of imposing a finite second moment using the $2^\text{nd}$-moment constrained mixture IS as in Section \ref{sec:statespace}.

The simulation study generates 100 time series of length $T=500$ from model \eqref{eq:poisson}.
Given the model parameter $\psi=(\beta,\phi,\sigma^2)$, for each time series,
we perform 100 likelihood evaluations at $\psi$ using $S=10^4$ importance samples for each likelihood evaluation.
Table \ref{tab:ss_model} reports the following performance measures averaged over the 100 realizations of the time series $\{y_t,t=1,...,T\}$
and 100 evaluations: (i) the ratio of the variance of the weights;
(ii) the ratio of the Monte Carlo standard errors (MCE) of the likelihood evaluations,
with the SPDK sampler that does not impose the existence condition as the benchmark in both cases;
(iii) the computing time per one likelihood evaluation;
and (iv) the proportion of the 100 replications in which the original SPDK sampler satisfies the criterion \eqref{eq:p.d.}
for the existence of the second moment.

\begin{table}[htbp]
\begin{center}
\begin{tabular}{lcc|cc}
\hline\hline
&\multicolumn{2}{c|}{$\beta=-1.4,\phi=0.99,\sigma^2=1$}&\multicolumn{2}{c}{$\beta=-1.4,\phi=0.8,\sigma^2=0.18$}\\
		& not imposing	&imposing& not imposing & imposing\\
\hline
Variance 	&1		&0.0005	&1		&0.77\\
MCE 		&1		&0.97	&1		&0.92\\	
CPU (second) 	&0.46    	&0.52	&0.21		&0.23 \\
Finite variance &0		&-	&0		&-\\
\hline\hline
\end{tabular}
\caption{Likelihood evaluation for the Poisson state space model.
The table reports the ratio of the variance of the weights,
the ratio of the Monte Carlo standard errors (MCE)
with the SPDK sampler without imposing the existence condition as the benchmark,
the CPU time per likelihood evaluation, and
the proportion of the 100 replications in which the original SPDK sampler satisfies the existence condition.
}
\label{tab:ss_model}
\end{center}
\end{table}

We estimate the likelihood for each simulated dataset at two different values of the model parameters $\psi$. The first parameter vector is $\psi=(-1.4,0.99,1)$, which represents an extreme case away from the DGP values. From Proposition \ref{prop:pseudo_var}, higher values of $\phi$ and $\sigma^2$ lead to a stricter restriction on the approximating linear state space model in order for the variance of the importance sampler to exist. In this case, imposing the existence condition leads to a substantial improvement in terms of both variance of the weights and the Monte Carlo errors of the likelihood evaluation. The original SPDK method always fails the criterion of Proposition \ref{prop:gauss_trans} for the existence of the second moment. In the second case, we evaluate the likelihood at the true parameter values. In this case imposing the existence condition does not improve much on the original SPDK method. This result holds despite the theoretical infinite variance of the SPDK importance sampler.

%-------------------------------------------------%
\subsection{Estimating a panel data model with autoregressive random effects}
%-------------------------------------------------%
This simulation study estimates a panel data model with autoregressive random effects.
We generate a data set from a Poisson longitudinal model
\begin{eqnarray}\label{Exa:Poisson}
y_{it}&\sim&\text{Poisson}(\lambda_{it}),\;\;\lambda_{it}=\exp(x_{it}'\beta+\alpha_{it}),\notag\\
\alpha_{i,t+1}&=&\phi\alpha_{it}+\eta_{it},\;\;\alpha_{i1}\sim N(0,\sigma_\alpha^2),\notag\\
\eta_{it}&\stackrel{iid}{\sim}&N(0,\sigma^2),\;\;\sigma_\alpha^2=\sigma^2/(1-\phi^2),\notag
\end{eqnarray}
with $t=1,...,T$ and $i=1,...,m$.
We set $m=20$, $\beta=(1.4, -1)'$, $\phi=0.8$,
$x_{it}=(1, z_{it})'$ with $z_{it}$ generated randomly from a uniform distribution on [0,1].
The unknown model parameters are $\psi=(\beta,\phi,\sigma^2)$.

%-----------------------------------------%
\subsubsection{Likelihood estimation}\label{subsec:llh_estimation}
%-----------------------------------------%
We investigate the potential benefit of imposing the condition
that the importance weights have a finite variance in the context of likelihood evaluation at a particular value of the model parameters.
As noted in Section \ref{sec:panel},
the likelihood of this panel data model decomposes into a product of lower dimensional integrals.
We estimate each of these integrals
by first using the SPDK method to obtain the importance parameters,
then construct the $2^\text{nd}$-moment constrained mixture IS as described in Section \ref{sec:mixture}.

For the reasons discussed in Section \ref{subsec:toy example}, we compare this
$2^\text{nd}$-IS to a $t$-IS with $\nu=5$ degrees of freedom, noting again
that using such a $t$ importance density does not guarantee the existence of the second moments of the weights.

\begin{table}[h]
\centering
\vskip2mm
{\small
\begin{tabular}{cccccc|ccc}
\hline\hline
	&		&		&\multicolumn{3}{c|}{Case 1}&\multicolumn{3}{c}{Case 2}\\
$T$	&$\sigma^2_\alpha$&Sampler	&Variance	&MCE	&CPU	&Variance	&MCE	&CPU\\
\hline
20	&0.5		&$t$-IS		&1		&1	&0.15	&1		&1	&0.17	\\
	&		&$2^\text{nd}$-IS&0.21		&0.48	&0.10	&0.55		&0.78	&0.10	\\
	&1		&$t$-IS		&1		&1	&0.15	&1		&1	&0.16	\\
	&		&$2^\text{nd}$-IS&0.32		&0.59	&0.10	&0.57		&0.79	&0.10	\\
\hline
50	&0.5		&$t$-IS		&1		&1	&0.22	&1		&1	&0.22	\\
	&		&$2^\text{nd}$-IS&0.23		&0.50	&0.18	&0.48		&0.76	&0.16	\\
	&1		&$t$-IS		&1		&1	&0.22	&1		&1	&0.22	\\
	&		&$2^\text{nd}$-IS&0.32		&0.65	&0.18	&0.55		&0.79	&0.16	\\
\hline
\hline
\end{tabular}
}
\caption{Likelihood estimation for the panel data model.
The table reports the ratio of the variance of weights and
the ratio of the Monte Carlo standard errors (MCE) of the likelihood estimates
with the $t$-IS as the benchmark.
CPU is the computing time per one likelihood evaluation.
}\label{t:llh estimation}
\end{table}

We draw 100 data sets from model \eqref{Exa:Poisson}
and perform 100 likelihood evaluations at a given model parameter vector $\psi$.
We consider two different sets of the model parameters:
in the first case $\psi$ is the vector of the true parameters that generate the data,
and in the second case $\psi$ is set to $(0,0,0,1)$ which is far from the true values.
We base each likelihood evaluation on $S=10^3$ importance samples.
Common random numbers (CRN) are used.
Similarly to Section \ref{subsec:non-Gau_ss}, Table \ref{t:llh estimation} reports
the ratio of the variance of the weights,
the ratio of the Monte Carlo standard errors (MCE) of the likelihood estimates
with the $t$-IS as the benchmark,
and computing time per one likelihood evaluation.
Note that each likelihood evaluation consists of estimating $m$ separate integrals,
so the variance reported here has been averaged over the $m$ panels.
The simulation results suggest that it is
beneficial to impose the condition that the importance weights have a finite variance.

Table \ref{t:llh estimation} also shows that although imposing the existence
of the second moments takes extra CPU time, the $2^\text{nd}$-IS method overall takes less time than the $t$-IS because working with a $t$ distribution takes more time than working with the constrained Gaussian mixture density.

\subsubsection{Bayesian inference}\label{subsubsec:Bayesian inference}
%-----------------------------------------%
This section explores the potential benefit of imposing the existence condition in the context of Bayesian inference.
Bayesian inference for the panel data model \eqref{Exa:Poisson} can be carried out
using the MCMC approach to sample from the posterior $p(\psi|y)$,
in which the likelihood used within the Metropolis-Hastings algorithm
is estimated unbiasedly by importance sampling.
We put a normal prior $N(0,\tau_0I_p)$ on $\beta$,
a reference prior $p(\phi)\propto1/\sqrt{1-\phi^2}$ on the correlation coefficient $\phi$ \citep{Berger:1994}
and $p(\sigma^2)\propto 1/\sigma^2$,
and set $\tau_0=100$.

%\begin{figure}[h]
%\centering
%\includegraphics[width=.8\textwidth,height=.4\textheight]{panel_example_figT20.eps}
%\caption{Panel data model example: Plots of the MCMC chains and autocorrelations for $\beta$, $\phi$
%and $\sigma^2$ for the case with $T=20$ and $\sigma^2_\alpha=0.5$.}
%\label{f:panel_example_figT20}
%\end{figure}

\begin{figure}[h]
\centering
\includegraphics[width=.8\textwidth,height=.4\textheight]{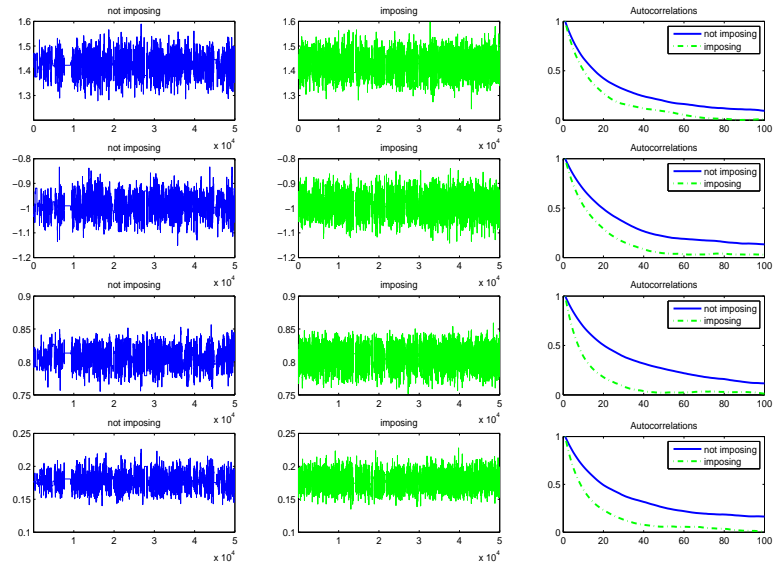}
\caption{Panel data model example. Plots of the MCMC chains and autocorrelations for $\beta$, $\phi$
and $\sigma^2$ for the case with $T=80$ and $\sigma^2_\alpha=1$.}
\label{f:panel_example_figT80}
\end{figure}

We consider two MCMC samplers.
The first sampler uses the $2^\text{nd}$-IS to estimate the likelihood to ensure the finite second moment of the importance weights,
and leads to a finite variance of the log likelihood estimator
which is important for MCMC simulation based on an estimated likelihood \citep{Pitt:2012b}.
The second MCMC sampler uses the $t$-IS as in Section \ref{subsec:llh_estimation} to estimate the likelihood.
Each sampler is run for 50,000 iterations after 50,000 burn-in iterations
using $S=200$ importance samples to estimate each likelihood.
The Markov chains are drawn by the adaptive random walk Metropolis-Hastings algorithm in \cite{Haario:2001}.

We use the integrated autocorrelation time (IACT)
and the acceptance rate of the Metropolis-Hastings algorithm as the performance measures.
For a scalar parameter $\theta$ the IACT is defined as \citep{Liu:2001}
\bqns
\text{IACT} = 1+2\sum_{j=1}^\infty\rho_j,
\eqns
where $\rho_j$ is the $j$th lag autocorrelation of the iterates of $\theta$ in the MCMC scheme.
We estimate the IACT by
\bqns
\widehat{\text{IACT}}=1+2\sum_{j=1}^{L^*}\widehat\rho_j,
\eqns
where $\widehat\rho_j$ is the $j$th lag sample autocorrelation
and $L^*=\min\{1000,L\}$,  with $L$ the first index $j$ such that $|\widehat\rho_j|\leq2/\sqrt{K}$
where $K$ is the sample size used to estimate $\widehat\rho_j$.
When summarizing the performance of the MCMC with $d$ parameters estimated, we
report, for simplicity, the average IACT over the $d$ parameters.

We investigate the performance of the two MCMC samplers for four combinations of $T$ and $\sigma^2$.
Figures \ref{f:panel_example_figT80} plot the iterates and the autocorrelations for
the four parameters $\beta$ (first two rows), $\phi$ (third) and $\sigma^2$ (last)
for the case with $\sigma^2_\alpha=1$ and $T=80$,
which shows that the sampler which imposes the existence condition works better.
Table \ref{t:IS-MCMC} summarizes the acceptance rates and the IACT ratios averaged over 5 replications. The table also reports the CPU time taken for each MCMC iteration.
The results suggest that the new method is beneficial in cases with large $\sigma^2$, after taking into account the higher computing time required to impose the variance existence condition.

\begin{table}[h]
\centering
\vskip2mm
{\small
\begin{tabular}{ccccccc}
\hline\hline
$T$	&$\sigma^2_\alpha$&Imposing	&Acc. rate (\%)	&IACT ratio& CPU (second)\\
\hline
20	&0.5		&No		&20		&1	&0.015\\
	&		&Yes		&25		&0.79	&0.033\\
	&1		&No		&6		&1	&0.016\\
	&		&Yes		&12		&0.34	&0.033\\
\hline
80	&0.5		&No		&20		&1	&0.064\\
	&		&Yes		&23		&0.95	&0.135\\
	&1		&No		&1.8		&1	&0.061\\
	&		&Yes		&4.8		&0.54	&0.132\\
\hline
\hline
\end{tabular}
}
\caption{
Bayesian inference for the panel data model.
The table reports the acceptance rates (Acc. rate), the integrated autocorrelation time (IACT) ratios
with the MCMC sampler that does not impose the existence condition as the benchmark,
and the CPU time taken for each MCMC iteration.
All values are averaged over 5 replications.
}\label{t:IS-MCMC}
\end{table}

%-----------------------------------------%
\subsection{An application to the anti-epileptic drug dataset}\label{Exa:GLMM}
%-----------------------------------------%
The anti-epileptic drug longitudinal dataset \citep[see, e.g.,][p.346]{Fitzmaurice:2011}
consists of seizures counts on 59 epileptic patients over 5 time-intervals of treatment.
The objective is to study the effects of the anti-epileptic drug on the patients.
Following \cite{Fitzmaurice:2011}, we consider the mixed effects Poisson regression model
\begin{eqnarray*}
p(y_{ij}|\beta,\alpha_i)&=&\text{Poisson}(\exp(\eta_{ij})),\\
\eta_{ij}&=&c_{ij}+\beta_1+\beta_2\text{time}_{ij}+\beta_3\text{treatment}_{ij}+\beta_4\text{time}_{ij}\times \text{treatment}_{ij}\\
&&+\alpha_{i1}+\alpha_{i2}\text{time}_{ij},
\end{eqnarray*}
$j=0,1,...,4$, $i=1,...,59$ and $c_{ij}$ is an offset.
As in \cite{Fitzmaurice:2011}, the offset $c_{ij}=\log(8)$ if $j=0$ and $c_{ij}=\log(2)$ for $j>0$,
$\text{time}_{ij}=j$, $\text{treatment}_{ij}=0$ if patient $i$ is in the placebo group
and $\text{treatment}_{ij}=1$ if in the treatment group.
The $\alpha_i=(\alpha_{i1},\alpha_{i2})'\sim N(0,Q^{-1})$ are random effects that need to be integrated out.
We consider a normal prior $N_p(0,\tau_0I_p)$ for $\beta$
and a Wishart $\mathcal W(\nu_0,\tau_0I_u)$ prior for the precision $Q$.
In order to have flat priors, we select $\tau_0=1000$, $\nu_0=u+1$.
In the MCMC scheme, we first use the Leonard and Hsu transformation
$Q = \exp(\Sigma)$,
where $\Sigma$ is an unconstrained symmetric matrix,
and reparameterize $Q$ by the lower-triangle elements $\theta_Q$ of $\Sigma$,
which is an one-to-one transformation between $Q$ and $\theta_Q$.
We then use the adaptive random walk Metropolis-Hastings algorithm in \cite{Haario:2001}
to sample from the posterior $p(\beta,\theta_Q|y)$.
The dimension of $\psi=(\beta,\theta_Q)$ is $4+2(2+1)/2=7$.

\begin{figure}[h]
\centering
\includegraphics[width=1\textwidth,height=.6\textheight]{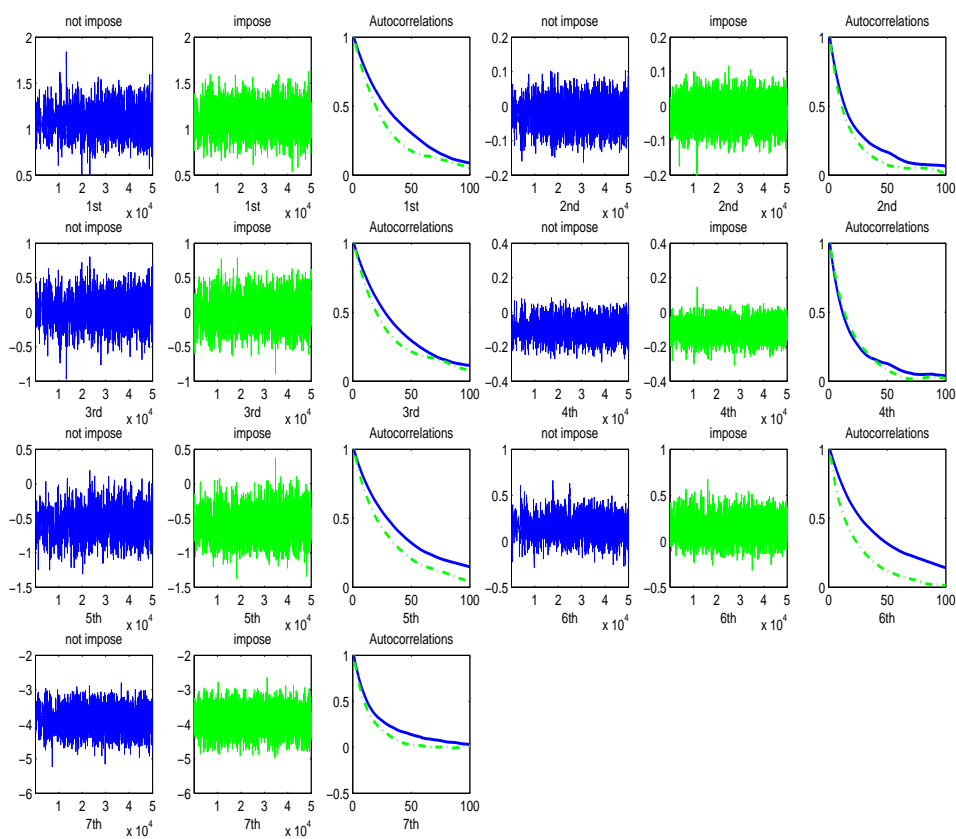}
\caption{Drug example: The figure plots the marginal chains and their corresponding first 100 autocorrelations,
generated by the MCMC sampler based on the $2^\text{nd}$-IS (green) and the MCMC sampler based on the $t$-IS (blue).
}
\label{f:drug_example}
\end{figure}

Similarly to Section \ref{subsubsec:Bayesian inference},
we run two MCMC samplers based on $2^\text{nd}$-IS and $t$-IS. 
Each sampler is run for 50000 iterations after discarding 50000 burn-in iterations.
The number of samples used in each likelihood estimation is $S=200$.
Figure \ref{f:drug_example} plots the iterates of the parameters
as well as the autocorrelations generated by the two samplers.
The IACT values of the samplers imposing and not imposing the existence condition
are 19.6 and 29.2.
That is, imposing the condition leads to a sampler
that is roughly 1.5 times more efficient than not imposing.
The acceptance rates averaged over replications of the sampler imposing and not imposing the existence condition
are 22.3\% and 18.8\%, respectively.
This suggests that it is beneficial to impose the finite second moment condition of the importance weight
within a MCMC sampler.
The two samplers respectively take 0.079 and 0.058 seconds to run each MCMC iteration.

\clearpage
\bibliographystyle{apalike}
\bibliography{references}

\newpage
\appendix
\section*{Appendix A}
\begin{proof}[Proof of Proposition \ref{prop_general}]
Let $\text{const}_1$, $\text{const}_2$,... be generic constants.
By equation \eqref{eq:bounding}
\begin{align*}
\log\omega(\alpha)&=l(\alpha)+\log p(\alpha)-\log g(\alpha)\\
&\leq\text{const}_1-\frac{1}{2}(\alpha-\xi)^{\prime}Q(\alpha-\xi)+\frac{1}{2}(\alpha-\mu^{\ast})^{\prime
}Q^{\ast}(\alpha-\mu^{\ast})\\
&  =\text{const}_2+\frac{1}{2}(\alpha-\mu^{+})^{\prime}Q^{+}(\alpha-\mu^{+}),
\end{align*}
where $Q^{+}=Q^{\ast}-Q$, $\mu^{+}=\left(  Q^{+}\right)  ^{-1}(Q^{\ast}\mu^{\ast}-Q\xi)$.
Hence
\[
n\log\omega(\alpha)\leq\text{const}_{3}+\frac{1}{2}(\alpha-\mu^{+})^{\prime}%
nQ^{+}(\alpha-\mu^{+}),
\]
and
\begin{align*}
E_{g}[\omega(\alpha)^{n}]  &  \leq\text{const}_{4}\times\int\exp\left\{  \frac{1}%
{2}(\alpha-\mu^{+})^{\prime}nQ^{+}(\alpha-\mu^{+})\right\}  g(\alpha)d\alpha\\
&  =\text{const}_{5}\times\int\exp\left\{-\frac{1}{2}(\alpha-a^{+})^{\prime}%
V(\alpha-a^{+})\right\}  d\alpha,
\end{align*}
where $V=Q^{\ast}-nQ^{+}=Q^{\ast}-n(Q^{\ast}-Q)$, $Va^{+} =Q^{\ast}\mu^{\ast}-nQ^{+}\mu^{+}$. Since $V=Q^{\ast}-n(Q^{\ast}-Q)>0$ by assumption, the integrand in the last equation corresponds to the kernel of a multivariate Gaussian density. The integral is therefore finite.
\end{proof}

\begin{proof}[Proof of Proposition \ref{prop:gauss_trans}]
Proof of (i) follows directly from Proposition \ref{prop_general}.
To prove (ii), let $V=Q^{\ast}-n(Q^{\ast}-Q)<0$.
\begin{align*}
\log\omega(\alpha)&=l(\alpha)+\log p(\alpha)-\log g(\alpha)\\
&=\text{const}_1+l(\alpha)-\frac{1}{2}(\alpha-\xi)^{\prime}Q(\alpha-\xi)+\frac{1}{2}(\alpha-\mu^{\ast})^{\prime
}Q^{\ast}(\alpha-\mu^{\ast})\\
& =\text{const}_2+l(\alpha)+\frac{1}{2}(\alpha-\mu^{+})^{\prime}Q^{+}(\alpha-\mu^{+}),
\end{align*}
where $Q^+=Q^*-Q$. So
\begin{align*}
E_{g}[\omega(\alpha)^{n}]&=\text{const}_{3}\times\int_{\mathbb{R}^d}\exp\left\{nl(\alpha)+\frac{1}%
{2}(\alpha-\mu^{+})^{\prime}nQ^{+}(\alpha-\mu^{+})\right\}  g(\alpha)d\alpha\\
&  =\text{const}_{4}\times\int_{\mathbb{R}^d}\exp\left\{nl(\alpha)+\zeta'\alpha-\frac{1}{2}\alpha^{\prime}%
V\alpha\right\}  d\alpha,
\end{align*}
where $\zeta$ is a constant vector.
We can write $\frac12\alpha'V\alpha=\frac12v_{jj}\alpha_j^2+B\alpha_j+C$,
where $v_{jj}<0$ as $V<0$ and $B$ and $C$ involve only $\alpha_{-j}=(\alpha_1,...,\alpha_{j-1},\alpha_{j+1},...,\alpha_{d})'$.
The moment $E_{g}[\omega(\alpha)^{n}]$ can now be written as
\begin{align*}
E_{g}[\omega(\alpha)^{n}]&=\text{const}_{4}\times\int_{\mathbb{R}^{d-1}}\left(\int_{\mathbb{R}}\exp\left\{nl(\alpha)+\zeta'\alpha-\frac12v_{jj}\alpha_j^2-B\alpha_j-C\right\}d\alpha_j\right)d\alpha_{-j}.
\end{align*}
By \eqref{eq:more assumption}, for any fixed $\alpha_{-j}$,
\bqns
nl(\alpha)+d'\alpha-\frac12v_{jj}\alpha_j^2-B\alpha_j-C=\alpha_j^2\left(-\frac12v_{jj}+\frac{nl(\alpha)+d'\alpha-B\alpha_j-C}{\alpha_j^2}\right)\longrightarrow+\infty
\eqns
when either $\alpha_j\to-\infty$ or $\alpha_j\to+\infty$.
It follows that the integral over $\alpha_j$ is infinite for any fixed $\alpha_{-j}$,
thus $E_{g}[\omega(\alpha)^{n}]$ is infinite.
%without loss of generality, assume that $\lim_{\alpha_1\to+\infty}(l(\alpha)-k_1-\delta_1\alpha_1)=0$.
%This implies that, for any $\epsilon>0$, there exists a number $M$ such that $nl(\alpha)>nk_1+n\delta_1\alpha_1-n\epsilon$
%for any $\alpha>M$. Hence
%\begin{align*}
%&\int\exp\left\{nl(\alpha)-\frac{1}{2}A\alpha_1^2-\frac12B\alpha_1-\frac12 C\right\} d\alpha_1>\\
%&\phantom{ccccc}\int_M^{+\infty}\exp\left\{nk_1+n\delta_1\alpha_1-n\epsilon-\frac{1}{2}A\alpha_1^2-\frac12B\alpha_1-\frac12 C\right\} d\alpha_1=\infty
%\end{align*}
%as $A<0$, which implies that $E_{g}[\omega(\alpha)^{n}]=\infty$.
\end{proof}

%------------------------------------%
\begin{proof}[Proof of Proposition \ref{prop_mixture}]
%------------------------------------%
Since $g(\alpha)\geq \pi g_1(\alpha)$, $\sup(g)\supseteq\sup(g_1)\supseteq\sup(f)$.
\bqns
\int_{\sup(g)}\left(  \frac{f(\alpha)}{g(\alpha)}\right)  ^{n}g(\alpha)d\alpha&=&\int_{\sup(g_1)}\left(  \frac{f(\alpha)}{g(\alpha)}\right)  ^{n}g(\alpha)d\alpha
+\int_{\sup(g)\setminus\sup(g_1)}\left(  \frac{f(\alpha)}{g(\alpha)}\right)  ^{n}g(\alpha)d\alpha\\
&=&A+B.
\eqns
We have
\bqns
A=\int_{\sup(g_1)}\left(  \frac{f(\alpha)}{g(\alpha)}\right)  ^{n}g(\alpha)d&=&\int_{\sup(g_1)}\left(  \frac{f(\alpha)}{g_1(\alpha)}\right)  ^{n}\left(  \frac{g_1(\alpha)}{g(\alpha)}\right)  ^{n-1}g_1(\alpha)d\alpha\\
&\leq&\frac{1}{\pi^{n-1}}\int_{\sup(g_1)}\left(  \frac{f(\alpha)}{g_{1}(\alpha)}\right)  ^{n}%
g_{1}(\alpha)d\alpha<\infty.
\eqns
If $\sup(g)\setminus\sup(g_1)=\emptyset$, $B=0$.
Otherwise, $f(\alpha)=0$ on $\sup(g)\setminus\sup(g_1)$, and therefore $B=0$.
This completes the proof.
\end{proof}

%------------------------------------%
\begin{proof}[Proof of Proposition \ref{prop:pseudo_var}]
%------------------------------------%
From Proposition \ref{prop:gauss_trans}, we need to have $Q-(n-1)C>0$.
This is equivalent to having all the determinants $\Lambda_t>0$, $t=1,...,T-1$ and $\wt\Lambda_T>0$,
where the $\Lambda_t$ and $\wt\Lambda_T$ are given in \eqref{eq:diff eq1} and \eqref{eq:diff eq2}, with $v_t=v$.
Let $c=1+\phi^{2}-\sigma^{2}(n-1)/v$. The general solution of the recursion equation \eqref{eq:diff eq1} is a linear combination of the powers of the roots of the characteristic polynomial
\begin{equation}\label{eq:cha poly}
r^{2}-cr+\phi^2=0
\end{equation}
if the roots are unequal, as we discuss bellow.

Consider first the case where $v>(n-1)\sigma^2_\alpha(1+|\phi|)/(1-|\phi|)=(n-1)\sigma^2/(1-|\phi|)^2$.
Then $c>2|\phi|$, which ensures that the characteristic polynomial \eqref{eq:cha poly} has
two distinct and positive roots $r_1 = {c}/{2}+\frac12\sqrt{c^2-4\phi^2}$ and $r_2 = {c}/{2}-\frac12\sqrt{c^2-4\phi^2}$.
Hence the general solution to the second-order difference equation \eqref{eq:diff eq1}
is
\bqns
\Lambda_t = Kr_1^t+Lr_2^t,\;\;t=0,...,T-1,
\eqns
where $K,L$ are constants determined by the boundary conditions $K+L=1$ and $Kr_1+Lr_2=\Lambda_1$.
Note that $v>(n-1)\sigma^2/(1-|\phi|)^2>(n-1)\sigma^2$, which implies that $\Lambda_1=1-(n-1)\sigma^2/v>0$.
Using the boundary conditions, we obtain
\[K=\frac{c-\phi^2-r_2}{r_1-r_2},\;\;L=\frac{r_1-c+\phi^2}{r_1-r_2}.\]
As $r_2<c/2$, $|\phi|<1$ and $r_1>r_2$, we have that $K>(c/2-\phi^2)/(r_1-r_2)>(|\phi|-\phi^2)/(r_1-r_2)>0$.
Therefore, for $t=0,...,T-1$,
\bqns
\Lambda_t = Kr_1^t+(1-K)r_2^t=K(r_1^t-r_2^t)+r_2^t>0.
\eqns
We now check that $\wt\Lambda_T>0$ as well, by observing that for any $t\geq 0$,
\begin{align*}
\Lambda_t=Kr_1^t+Lr_2^t&=\frac{c-\phi^2-r_2}{r_1-r_2}r_1^t+\frac{r_1-c+\phi^2}{r_1-r_2}r_2^t\\
&=\frac{cr_1^t-\phi^2r_1^t-\phi^2r_1^{t-1}}{r_1-r_2}+\frac{\phi^2r_2^{t-1}-cr_2^t+\phi^2r_2^t}{r_1-r_2}\\
&=\frac{r_1^{t-1}(cr_1-\phi^2)-\phi^2r_1^t}{r_1-r_2}+\frac{r_2^{t-1}(\phi^2-cr_2)+\phi^2r_2^t}{r_1-r_2}\\
&=\frac{r_1^{t+1}-\phi^2r_1^t}{r_1-r_2}+\frac{-r_2^{t+1}+\phi^2r_2^t}{r_1-r_2}\\
&=\frac{r_1^{t+1}-r_2^{t+1}-\phi^2(r_1^t-r_2^t)}{r_1-r_2}.
\end{align*}
In the above, we use the fact that $r_1r_2=\phi^2$ and $cr_i-\phi^2=r_i^2$, $i=1,2$.
Hence,
\begin{align*}
\wt\Lambda_T&=\left(\{1+\phi^2-(n-1)\sigma^2/v\}\Lambda_{T-1}-\phi^2\Lambda_{T-2}\right)-\phi^2\Lambda_{T-1}\\
&=\Lambda_T-\phi^2\Lambda_{T-1}\\
&=\frac{r_1^{T+1}-r_2^{T+1}-\phi^2(r_1^T-r_2^T)}{r_1-r_2}-\phi^2\frac{r_1^{T}-r_2^{T}-\phi^2(r_1^{T-1}-r_2^{T-1})}{r_1-r_2}\\
&=\frac{r_1^{T-1}(r_1-\phi^2)^2-r_2^{T-1}(r_2-\phi^2)^2}{r_1-r_2}.
\end{align*}
It is easy to check that $-(r_1-\phi^2)<r_2-\phi^2<r_1-\phi^2$, thus $(r_2-\phi^2)^2<(r_1-\phi^2)^2$.
This, together with $r_1>r_2$, implies that $\wt\Lambda_T>0$.

Consider the case where $v=(n-1)\sigma^2_\alpha(1+|\phi|)/(1-|\phi|)=(n-1)\sigma^2/(1-|\phi|)^2$.
The characteristic polynomial \eqref{eq:cha poly} has two equal roots $r_1=r_2=r=|\phi|$.
Then the solution to the difference equation \eqref{eq:diff eq1} is $\Lambda_t=(A_0+A_1t)r^t$
with $A_0=1$ and $A_1=1-|\phi|$.
If $\phi\not=0$, it is obvious that $\Lambda_t>0$ for $t=0,...,T-1$,
and for $t=T$,
\begin{align*}
\wt\Lambda_T=\Lambda_T-\phi^2\Lambda_{T-1}&=(A_0+A_1T)r^T-\phi^2\big((A_0+A_1(T-1))r^{T-1}\big)\\
&=|\phi|^T(1-|\phi|)(1+|\phi|+T(1-|\phi|))>0.
\end{align*}
This completes the proof.
\end{proof}

%%------------------------------------%
%\begin{proof}[Proof of Proposition \ref{prop:pseudo_var}]
%%------------------------------------%
%From Proposition \ref{prop:gauss_trans}, we need to have $Q-(n-1)C>0$ where
%the diagonal matrix $C_{tt}=-1/v.$ This matrix has the tridiagonal form of
%(\ref{eq:matrix_form}). The (top left $t\times t$) determinants $\Lambda_{t}$
%are all required to be positive, by Sylvester's criterion. From the structure of
%(\ref{eq:matrix_form}), it can be seen that the determinants satisfy,
%\[
%\Lambda_{t}-\{1+\phi^{2}-\sigma^{2}(n-1)/v\}\Lambda_{t-1}+\phi^{2}%
%\Lambda_{t-2}=0,
%\]
%for $t=2,3,...,T-1$ where $\Lambda_{0}=1,$ $\Lambda_{1}=1-\sigma^{2}/v$. In
%this case we need to investigate the roots of the above difference equation.
%The roots are given by the roots of the quadratic%
%\[
%r^{2}-c_{1}r-c_{2}=0,
%\]
%where $c_{1}=\{1+\phi^{2}-\sigma^{2}(n-1)/v\}$ and $c_{2}=-\phi^{2}$. Hence
%the roots are
%\[
%r=\frac{c_{1}}{2}\left\{  1\pm\sqrt{1-4\left(  \frac{\phi}{1+\phi^{2}%
%-\sigma^{2}(n-1)/v}\right)  ^{2}}\right\}  .
%\]
%Hence if $c_{1}>0$ so that $v>(n-1)\sigma^{2}\left(  1+\phi^{2}\right)  ^{-1}$
%the roots will be real, positive and distinct provided,%
%\[
%1+\phi^{2}-\sigma^{2}(n-1)/v>2\phi
%\]
%which implies
%\[
%v>(n-1)\sigma_{\alpha}^{2}\frac{(1+\phi)}{(1-\phi)},
%\]
%where $\sigma_{\alpha}^{2}=\sigma^{2}/(1-\phi^{2})$, the marginal variance of
%$\alpha_{t}.$
%\end{proof}

\section*{Appendix B}
\cite{SP97} and \cite{DK97} construct an importance sampler based on the following approximation of the target density
\begin{align*}
\log p(y|\alpha)+\log p(\alpha)  &  =l(\alpha)-\frac{1}{2}(\alpha-\mu
)^{T}Q(\alpha-\mu)\label{eqn:approx1}\\
&  \simeq l(\widehat{\alpha})+(\alpha-\widehat{\alpha})^{T}l^{\prime
}(\widehat{\alpha})-\frac{1}{2}(\alpha-\widehat{\alpha})^{T}C(\alpha
-\widehat{\alpha})-\frac{1}{2}(\alpha-\mu)^{T}Q(\alpha-\mu),\nonumber
\end{align*}
where $\widehat{\alpha}$ is the mode of $p(y|\alpha)p(\alpha)$ and $C$ is minus the Hessian of $l(\alpha)$ at $\widehat{\alpha}$.
In the framework of equation \eqref{idensity}, $B=l^{\prime}(\widehat{\alpha})+C\widehat{\alpha}$.

We can find the mode of $p(y|\alpha)p(\alpha)$ by using the following iterative algorithm, which is equivalent to a Newton-Raphson procedure for finding the maximum of the target density as a function of $\alpha$.

\bigskip

\noindent{\bf Algorithm: Constructing the SPDK importance density}
\begin{enumerate}
\item Start with an initial guess $\widehat{\alpha}$.
\item {\bf While} convergence criterion is not met {\bf do}
\begin{itemize}
  \item Calculate the matrix of second derivatives $C\gets-\left.  \frac{\partial^{2}l(\alpha)}{\partial\alpha\partial\alpha^{\prime}
}\right\vert _{\alpha=\widehat{\alpha}}$
  \item Update the other importance parameter $B\gets l^{\prime}(\widehat{\alpha})+C\widehat{\alpha}$.
  \item Update the covariance matrix $Q^{\ast-1}\gets (C+Q)^{-1}$.
  \item Update the estimate of the mode $\widehat{\alpha}\gets Q^{\ast-1}(B+Q\mu)$.
\end{itemize}
\item Set the mean of the importance density as $\mu^\ast=\widehat{\alpha}$.
\end{enumerate}

%\begin{algorithm}%\caption{Constructing the SPDK importance density (general case)}
%{Constructing the SPDK importance density (general case)}
%\begin{algorithmic}[1]
%\State Start with an initial guess $\widehat{\alpha}$.
%    \While{convergence criterion is not met}
%    \State Calculate the matrix of second derivatives. $C\gets-\left.  \frac{\partial^{2}l(\alpha)}{\partial\alpha\partial\alpha^{\prime}%
%}\right\vert _{\alpha=\widehat{\alpha}}$
%      \State Update the other importance parameter. $B\gets l^{\prime}(\widehat{\alpha})+C\widehat{\alpha}$.
%      \State Update the covariance matrix $Q^{\ast-1}\gets (C+Q)^{-1}$.
%      \State Update the estimate of the mode $\widehat{\alpha}\gets Q^{\ast-1}(B+Q\mu)$.
%    \EndWhile
%\State Set the mean of the importance density as $\mu^\ast=\widehat{\alpha}$.
%\end{algorithmic}
%\end{algorithm}

\cite{SP97} and \cite{DK97} show how to carry out the above computations efficiently using the Kalman filter and smoother for the state space model of Section \ref{sec:statespace}. In this case, it is unnecessary to calculate the items in Algorithm 1 explicitly, except for the mode. For this class of models, we can also obtain near optimal importance parameters $\left\{ b_t, C_t \right\}$ for $t=1,\ldots,T$ by using the NAIS method of \cite{NAIS}. The method is based on recursively minimising
\begin{equation}\label{proxprobt}
\underset{\chi _t}{\min} \int \lambda ^2 (\alpha_t, y_t;\psi) g(\alpha _t |y ;\psi) \dd\alpha_t
\end{equation}
where
\begin{equation}\label{lambdadef}
\lambda (\alpha _t, y_t;\psi) = \log p(y_t | \alpha_t; \psi) - a_t - b_t^{\prime} \, \alpha_t+\frac 12 \alpha_t^{\prime}C_t\alpha_t,
\end{equation}
for $t=1,\ldots,T$, given the current estimates of the optimal importance parameters. We obtain $g(\alpha _t |y ;\psi)$ by the Kalman filter and smoother and evaluate the integral \eqref{proxprobt} by numerical integration (in the univariate state case). Due to the linearity of $\lambda (\theta _t, y_t;\psi)$, the minimisation consists of a weighted OLS regression.

\section*{Appendix C}
We make $\wt\lambda_j$ continuous in $\lambda_j$ by defining
\begin{equation}\label{eq:impose2}
\wt\lambda_j=
\begin{cases}
\lambda_j,&\text{if}\;\;\lambda_j<\frac{1-\epsilon}{n-1},\\
\frac{1-\epsilon}{n-1},&\text{if}\;\;\lambda_j\geq\frac{1-\epsilon}{n-1}.
\end{cases}
\end{equation}
Then we still have that $nQ-(n-1)\wt Q=AV(I-(n-1)\wt\Lambda)V'A'>0$ because $\wt\lambda_j<1/(n-1)$ for all $j$.
However, the $\wt\lambda_j$ in \eqref{eq:impose2} is not a differentiable function of $\lambda_j$.
To make $\wt Q $ continuous and differentiable in $\psi$, we proceed as follows.
Let $\delta>0$ be some small number and denote $\tau=(1-\epsilon)/(n-1)$.
We define
\begin{equation}\label{eq:impose3}
\wt\lambda_j=
\begin{cases}
\lambda_j,&\text{if}\;\;\lambda_j<\tau-\delta,\\
f(\lambda_j),&\text{if}\;\;\tau-\delta\leq\lambda_j<\tau,\\
\tau,&\text{if}\;\;\lambda_j\geq\tau,
\end{cases}
\end{equation}
where $f(\lambda_j)=a\lambda_j^3+b\lambda_j^2+c\lambda_j+d$ satisfying
$f(\tau-\delta)=\tau-\delta$, $f(\tau)= \tau$, $\dot{f}(\tau-\delta) = 1$ and $\ddot{f}(\tau) = 0$.
Then $\wt\lambda_j$ defined in \eqref{eq:impose3} is continuous and differentiable in $\lambda_j$.
After some algebra, we obtain $a=-1/\delta^2$, $b=3\tau/\delta^2-2/\delta$, $c=4\tau/\delta-3\tau^2/\delta^2$ and $d=\tau+\tau^3/\delta^2-2\tau^2/\delta$.
Now, $h(\lambda_j)=f'(\lambda_j)=3a\lambda_j^2+2b\lambda_j+c$ is an open-down parabola with $h(\tau-\delta)=1$ and  $h(\tau)=0$,
which implies that $h(\lambda_j)\geq0$ for all $\lambda_j$ in the interval $(\tau-\delta,\tau)$.
It follows that $f(\lambda_j)\leq f(\tau)=\tau$ for all $\lambda_j\in(\tau-\delta,\tau)$.
We have proved that $\wt\lambda_j\leq\tau<1/(n-1)$,
which ensures that $nQ-(n-1)\wt Q=AV(I-(n-1)\wt\Lambda)V'A'>0$ as required.
In our implementation we take $\delta=\epsilon=10^{-5}$.

\section*{Appendix D}
Recall that the latent $\alpha_t$ follows an AR(1) process
$\alpha_{t+1} = d+\Phi\alpha_t+\eta_t$, $t=1,...,T-1$ with $\alpha_1\sim N(\mu,\Sigma_\alpha)$ and $\eta_t\sim N(0,\Sigma)$.
For any $\tau\geq 1$, let
\bqns
\Gamma(\tau)&=\text{cov}(\alpha_{t+1},\alpha_{t+1-\tau})=\Phi\text{cov}(\alpha_{t},\alpha_{t+1-\tau})=\Phi\Gamma(\tau-1).
\eqns
For $\tau=0$, $\Sigma_\alpha=\Gamma(0)$ satisfies the following equation
\begin{equation}\label{eq:Phi_0}
\Sigma_\alpha=\text{cov}(\alpha_{t+1},\alpha_{t+1})=\Phi\text{cov}(\alpha_{t},\alpha_{t})\Phi'+\Sigma=\Phi\Sigma_\alpha\Phi'+\Sigma.
\end{equation}
The covariance matrix of $\alpha=(\alpha_1',...,\alpha_T')'$ is
\bqns
Q^{-1}=\begin{pmatrix}
\Sigma_\alpha&(\Phi\Sigma_\alpha)'&...&(\Phi^{T-1}\Sigma_\alpha)'\\
\Phi\Sigma_\alpha&\Sigma_\alpha&...&(\Phi^{T-2}\Sigma_\alpha)'\\
\vdots&\vdots&\ddots&\vdots\\
\Phi^{T-1}\Sigma_\alpha&\Phi^{T-2}\Sigma_\alpha&...&\Sigma_\alpha
\end{pmatrix}.
\eqns
It can be shown that the inverse of this matrix is
\bqns%\label{eq:Qinverse}
Q=\begin{pmatrix}
\Sigma_\alpha^{-1}(I+\Phi(I-\Phi^2)^{-1}\Phi)&-\wt\Phi&0&...&0&0\\
-\wt\Phi&\wt\Gamma&-\wt\Phi&...&0&0\\
\vdots&\vdots&\vdots&&\vdots&\vdots\\
0&0&0&...&\wt\Gamma&-\wt\Phi\\
0&0&0&...&-\wt\Phi&\Sigma_\alpha^{-1}(I-\Phi^2)^{-1}
\end{pmatrix},
\eqns
where $\wt\Phi=\Sigma_\alpha^{-1}\Phi(I-\Phi^2)^{-1}$ and $\wt\Gamma=\Sigma_\alpha^{-1}(I-\Phi^2)^{-1}+\Sigma_\alpha^{-1}\Phi(I-\Phi^2)^{-1}\Phi$.

If $\alpha_t$ is scalar as in model \eqref{eq:ar1_state},
then using the notation in Section \ref{subsecsec:Sylvester},
$\Phi,\Sigma_\alpha,\Sigma$ become $\phi,\sigma_\alpha^2,\sigma^2$ respectively.
Solving $\Sigma_\alpha$ in \eqref{eq:Phi_0} gives $\sigma_\alpha^2=\sigma^2/(1-\phi^2)$
and
\bqns
Q=\begin{pmatrix}
1/\sigma^2&-\phi/\sigma^2&0&...&0&0\\
-\phi/\sigma^2&(1+\phi^2)/\sigma^2&-\phi/\sigma^2&...&0&0\\
\vdots&\vdots&\vdots&&\vdots&\vdots\\
0&0&0&...&(1+\phi^2)/\sigma^2&-\phi/\sigma^2\\
0&0&0&...&-\phi/\sigma^2&1/\sigma^2
\end{pmatrix}.
\eqns

\end{document}